\documentclass[10pt,twocolumn,nofootinbib,superscriptaddress]{revtex4-1}
\usepackage{amsfonts}
\usepackage{amsmath}
\usepackage{amssymb}
\usepackage{graphicx}%
\usepackage{comment}
\usepackage{color}
\usepackage{cancel}
\usepackage{ulem}
\usepackage[tiny]{titlesec}

\def\filetype{eps}

\begin{document}
\title{Competing Orders in a Dipolar Bose-Fermi Mixture on a Square Optical Lattice: Mean-Field Perspective}
\author{Jasen A.\ Scaramazza}
\affiliation{Department of Physics and Astronomy, Rowan University, Glassboro, New
Jersey 08028, USA}
\affiliation{Department of Physics and Astronomy, Rutgers University, Piscataway, New Jersey
08854, USA}
\author{Ben Kain}
\affiliation{Department of Physics, College of the Holy Cross, Worcester, Massachusetts 01610, USA}
\author{Hong Y.\ Ling}
\affiliation{Department of Physics and Astronomy, Rowan University, Glassboro, New
Jersey 08028, USA}
\affiliation{Kavli Institute for Theoretical Physics, University of California, Santa
Barbara, California 93106, USA }
\affiliation{ITAMP, Harvard-Smithsonian Center for Astrophysics, Cambridge, Massachusetts 02138, USA}


\begin{abstract}
\noindent We consider a mixture of a two-component Fermi gas and a single-component
dipolar Bose gas in a square optical lattice and reduce it into an effective
Fermi system where the Fermi-Fermi interaction includes the attractive
interaction induced by the phonons of a uniform dipolar Bose-Einstein
condensate. \ Focusing on this effective Fermi system in the parameter regime
that preserves the symmetry of $D_{4}$, the point group of a square, we
explore, within the Hartree-Fock-Bogoliubov mean-field theory, the phase
competition among density wave orderings and superfluid pairings.\ We
construct the matrix representation of the linearized gap equation in the
irreducible representations of $D_{4}$. \ We show that in the weak coupling
regime, each matrix element, which is a four-dimensional (4D) integral in
momentum space, can be put in a separable form involving a 1D integral, which
is only a function of temperature and the chemical potential, and a
pairing-specific \textquotedblleft effective\textquotedblright\ interaction,
which is an analytical function of the parameters that characterize the
Fermi-Fermi interactions in our system. \ We analyze the critical temperatures
of various competing orders as functions of different system parameters in
both the absence and presence of the dipolar interaction. We find that close
to half filling, the $d_{x^{2}-y^{2}}$-wave pairing with a critical
temperature in the order of a fraction of Fermi energy (at half filling) may
dominate all other phases, and at a higher filling factor, the $p$-wave
pairing with a critical temperature in the order of a hundredth of Fermi
energy may emerge as a winner. \ We find that tuning a dipolar interaction can
dramatically enhance the pairings with $d_{xy}$- and $g$-wave symmetries but
not enough for them to dominate other competing phases.

\end{abstract}
\maketitle


\section{Introduction}

Low-temperature crystal solids are full of wonderful and surprising phenomena,
that are of both fundamental and practical importance but often difficult to
observe owing to the imperfect nature and inflexibility of solid state
systems. \ Free of these shortcomings, cold atoms, offer a
pristine and highly controllable environment for creating novel quantum
matter. \ Experimental realization of the superfluid--Mott insulator
transition, a quintessential condensed matter phenomenon, in cold atom
lattices \cite{greiner02Nature.415.39} as exactly predicted by theory
\cite{fisher89PhysRevB.40.546,jaksch98PhysRevLett.81.3108} serves as one\ of
many remarkable examples. \ It is experimental developments like this,
along with the ability to create cold atom lattices that meet virtually any
need, that has motivated tremendous activity in the past decade in using cold
atom lattice models as an excellent platform for exploring and mimicking
complicated condensed matter phenomena \cite{bloch08RevModPhys.80.885}.

A subject of profound importance in condensed matter physics is
superconductivity, a coherent phenomenon arising from condensation of
correlated electron pairs, also known as Bardeen-Cooper-Schriffer (BCS) pairs,
on a macroscopic scale. \ A BCS pair, however, is not energetically favored unless there
exists an attraction between its constituents, which appears impossible
in an electron system where Coulomb repulsion dominates. \ The effort to solve
this mystery led, through the insights of Fl\"{o}lich, to the phonons of ionic
crystal lattice oscillations and the eventual understanding that the
interaction they induce between electrons is the source of the attraction
responsible for superconductivity in most materials \cite{schrieffer64Book}.

In cold atom physics, an analogous situation can arise from a Bose-Fermi
mixture, where phonons can exist as low energy excitations of a Bose-Einstein
condensate (BEC), describing the density fluctuation of the BEC
\cite{bijlsma00PhysRevA.61.053601}. The same phonons in the Bose gas can then
induce an attractive interaction between fermions in the Fermi gas
\cite{bijlsma00PhysRevA.61.053601}. \ This, along with the unprecedented
ability to tune parameters in such systems, e.g., two-body interaction and
dimensionality, allows for the possibility of clean and controllable
realization (or quantum simulation) of superfluid pairings in cold atom
Bose-Fermi mixtures
\cite{viverit02PhysRevA.66.023605,wang05PhysRevA.72.051604}.

Remarkable experimental progress has been made in creating quantum-degenerate
Bose-Fermi mixtures, e.g., $^{7}$Li-$^{6}$Li
\cite{truscott01Science.291.2570,schreck01PhysRevLett.87.080403}, $^{23}%
$Na-$^{6}$Li \cite{hadzibabic02PhysRevLett.88.160401}, $^{87}$Rb-$^{40}$K
\cite{ferrari02PhysRevLett.89.053202,roati02PhysRevLett.89.150403,inouye04PhysRevLett.93.183201,ferlaino06PhysRevA.73.040702}%
, $^{6}$Li-$^{87}$Rb \cite{deh08PhysRevA.77.010701}, and $^{23}$Na-$^{40}$K
\cite{park12PhysRevA.85.051602}. Most recently, a mixture where both
bosons and fermions are in the superfluid state has been realized
\cite{ferrier-Barbut14Science.345.1035}.  These developments together with the
recent upsurge of experimental efforts in achieving dipolar quantum gases, in
systems with $^{40}$K$^{87}$Rb
\cite{stuhler05PhysRevLett.95.150406,ospelkaus08NaturePhys.4.622,ni08Science.322.5899}%
, $^{6}$Li$^{40}$K
\cite{taglieber08PhysRevLett.100.010401,voigt09PhysRevLett.102.020405,pal14DAMOPSessionB3.8}%
, and\ $^{23}$Na$^{40}$K \cite{park14DAMOPSessionB3.6} heteronuclear
molecules, $^{87}$Rb spinor condensates
\cite{vengalattore08PhysRevLett.100.170403}, and $^{32}$Cr
\cite{ospelkaus09FaradayDiscuss142.351}, $^{164}$Dy
\cite{lu11PhysRevLett.107.190401} and $^{168}$Er
\cite{aikawa12PhysRevLett.108.210401} atoms, has opened up the exciting
possibility of creating dipolar Bose-Fermi mixtures with intriguing and unique
properties. Of particular relevance to the present work is a Bose-Fermi
mixture involving a two-component Fermi gas, examples of which include the
mixture of bosonic $^{41}$K with two fermionic species $^{40}$K$\;$and $^{6}%
$Li \cite{wu11PhysRevA.84.011601}, Feshbach molecules with two lowest
hyperfine states of the same fermionic species $^{6}$Li
\cite{shin08PhysRevLett.101.070404}, and ground state $^{6}$Li$^{40}$K
molecules with two fermionic species $^{6}$Li$~$and $^{40}$K
\cite{taglieber08PhysRevLett.100.010401,voigt09PhysRevLett.102.020405,pal14DAMOPSessionB3.8}%
. \ The last one is particularly motivating since the ground $^{6}$Li$^{40}$K
molecules are bosons with a large dipole moment of 3.6 Debye. \ In a dipolar
BEC \cite{santos00PhysRevLett.85.1791,yi00PhysRevA.61.041604}, the bosons
interact not only via the short-range s-wave but also the long-range
dipole-dipole interaction, and the dipolar interaction constitutes an
additional control knob inaccessible to a nondipolar condensate and may thus
open up new possibilities for engineering quantum gases with novel properties.

\begin{figure}
[ptb]
\begin{center}
\includegraphics[
width=2.5in
]%
{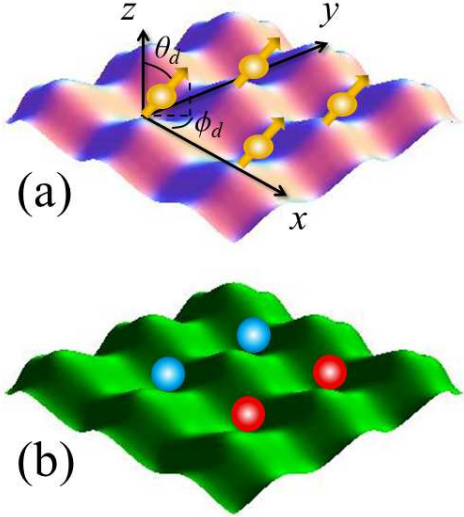}%
\caption{(Color online) A dipolar condensate (a) is mixed with a two-component
nondipolar Fermi gas (b) to form a dipolar Bose-Fermi mixture in a 2D square
optical lattice. All dipoles are pinned, by an external field, to $\left(
\theta_{d},\phi_{d}\right)  $, which are respectively the polar and azimuthal
coordinates defined relative to the 2D optical lattice.}%
\label{Fig:mixtureInLattice}%
\end{center}
\end{figure}

Motivated by this new prospect, we consider a dipolar Bose-Fermi mixture
loaded in a quasi two-dimensional (2D) trap, forming a square optical lattice,
where a two-component Fermi gas [Fig. \ref{Fig:mixtureInLattice}(b)] is mixed
with a single-component dipolar BEC [Fig. \ref{Fig:mixtureInLattice}(a)]. \ A
plethora of interesting phenomena have been predicted to occur in Bose-Fermi
mixtures in square lattices, including quantum phases of composite particles
involving pairing of fermions with boson particles or boson holes
\cite{lewenstein04PhysRevLett.92.050401}, superfluid-Mott insulator and
metal-insulator transition \cite{sengupta07PhysRevA.75.063625} and coexistent
phases \cite{sinha09PhysRevB.79.115124} within a slave-rotor mean-field
approximation, competition between phase separation and supersolid states
driven by a combination of van Hove instabilities and Fermi surface nesting
\cite{buchler03PhysRevLett.91.130404,buchler04PhysRevA.69.063603,orth09PhysRevA.80.023624}%
, superfluid pairings \cite{lim10PhysRevA.82.013616} and superfluid-Mott
insulator transition \cite{lim10PhysRevA.81.023404} in the presence of
artificial staggered magnetic field, and an interplay between density waves
and superfluids with unconventional pairing symmetries based on both a
mean-field study \cite{wang05PhysRevA.72.051604} and functional
renormalization group analysis
\cite{mathey06PhysRevLett.97.030601,klironomos07PhysRevLett.99.100401,huang13PhysRevB.88.054504}%
. \ However, thus far these studies have considered nondipolar bosons that
interact only through the repulsive hard-core potentials.

The focus of the present work is on the last topic --- competition between
density waves and unconventional superfluids, which are the cold atom analogs
of the unconventional superconductors in electron systems. \ Conventional
superconductors consist of spin-singlet s-wave BCS pairs
\cite{bardeen57PhysRev.108.1175} where both spin and orbital angular momenta
vanish, and it has been well established that the conventional (low-$T_{c}$)
superconductors are those where the pairing mechanism is dominated by the
phonon-induced electron-electron interaction. \ \ The study of unconventional
pairing states, where the gap parameters possess symmetries different from the
usual s-wave symmetry, began with the work by Anderson and Moral
\cite{anderson61PhysRev.123.1911}\ and that by Balian and Werthamer
\cite{balian63PhysRev.131.1553}, which led to the discovery of A- and B-phase
with $p$-wave symmetries in superfluid $^{3}$He
\cite{osheroff72PhysRevLett.28.885}. \ It has remained to this day an active
area of research, in large part, because of the discovery of high-$T_{c}$
superconductors in cuprate compounds by\ Bednorz and M\"{u}ller in 1986
\cite{bednorz86ZPhysB.64.189}, which are believed to possess unconventional
order parameters with $d_{x^{2}-y^{2}}$-wave symmetry
\cite{scalapino95PhysRep.250.329,tsuei00RevModPhys.72.969}.

In this paper, for simplicity, we shall restrict our attention to superfluids
where the spatial symmetry of a BCS pair can be classified according to the
irreducible representations of $D_{4}$, the point group of a square. \ An
important early work by Micnas \textit{et al. }%
\cite{micnas88PhysRevB.37.9410,micnas90RevModPhys.62.113} investigated, within
mean-field theory, superconducting phases of the same point group, based on an
extended Hubbard model with on-site repulsive and intersite attractive
interactions. It was shown explicitly that when the intersite interaction is
restricted to nearest neighbors, symmetrized two-body interactions in momentum
space can be cast into separable forms that support parings with $s$-,
$d_{x^{2}-y^{2}}$- and $p$-wave symmetries but not $d_{xy}$-and $g$-wave
symmetries. \ In a Bose-Fermi mixture, however, the effective Fermi-Fermi
interaction, which is mediated by the phonons of a BEC, has an interaction
range that depends on, among other things, the effective healing length. \ This
healing length is tunable by changing, e.g., the boson hopping amplitude. \ In
principle, as the healing length increases, the effective Fermi-Fermi
interaction may reach over many lattice sites
\cite{klironomos07PhysRevLett.99.100401}. \ Thus, classification by Micnas
\textit{et al. }\cite{micnas88PhysRevB.37.9410} holds only in the limit where
the healing length is much smaller than the lattice constant, which is
precisely the argument used by Bukov and Pollet in their recent work
\cite{bukov14PhysRevB.89.094502} to limit superfluids to
those with $s$-, $d_{x^{2}-y^{2}}$-, and $p$-wave symmetries. In contrast, we
do not place such a constraint on our model and we therefore include
superfluids with $d_{xy}$- and $g$-wave symmetries, which cannot be excluded a
priori in the presence of a long-range interaction \cite{balents13}. The inclusion of these two phases
allows us to take into consideration a wider choice of possibilities when
analyzing phase diagrams, making our studies, to some extent, more rigorous than those without them.

This paper is organized as follows. \ In Sec.\ II, we introduce a set of 2D
two-body interactions including onsite and offsite (nearest-neighbors) dipolar
ones, reducing our 3D trap model to a quasi 2D system in a square optical
lattice and describe it with the single-band Bose-Fermi Hamiltonian in both
real and reciprocal lattice space. \ 

In Sec.\ III, we introduce phonons that
obey the dispersion spectra of a homogeneous dipolar BEC. In the same section,
we eliminate the phonon degrees of freedom and arrive at an effective
Hamiltonian where fermions interact not only with the hard-core potential but
also the phonon-induced interaction. 

In Sec.\ IV, we take the Hartree-Fock Bogoliubov mean-field approach and
construct, in the space spanned by the base functions of the irreducible
representations of $D_{4}$, the matrix representation of the linearized gap
equation. The main complexity in this endeavor comes from evaluating the
related matrix elements which are 4D integrals in momentum space involving
symmetrized interactions that depend on momenta in a nontrivial way, something
absent in the simplest BCS theory. The gap equation in the simplest
BCS theory (for the s-wave superfluid pairing in continuous 3D models) can be
reduced to Eq. (\ref{3D S}) involving a 1D integral, which is the origin of
the well-known formula, Eq. (\ref{3D critical temperature}), that one uses to
estimate the critical temperature. A major contribution of the present work is
that by using a combination of algebra involving elliptical integrals and
symmetry considerations, we are able to reduce the 4D integrals in our model
into 1D integrals. \ This reduction results in a set of 1D integral equations
analogous to Eq. (\ref{3D S}) for all the possible superfluid phases
classified under the irreducible representations of group $D_4$. An example
derivation is given in the Appendix to highlight the main techniques we
employed to achieve such simplifications.

In Sec.\ V, we derive the conditions for the onset of the spin-density and
charge-density waves, two possible orderings arising from the instability of
the Fermi gas against the density-density fluctuation. \ 

In Sec. VI, we apply the formulas that we developed in Sec. IV
and V and perform a detailed investigation of the effect that various system
parameters (e.g. the healing length, the Fermi filling factor, and the dipolar
interaction) have on the critical temperatures for the onset of various
competing orders.

Finally, in Sec.\ VII, we summarize the results and highlight the significance of our work.

\section{Model and Hamiltonians for the Dipolar Bose-Fermi Mixture}

Let us now describe our model in some detail. \ The mixture under
consideration consists of a single-component BEC made up of dipolar bosons
of mass $m_{b}$ [Fig. \ref{Fig:mixtureInLattice}(a)] and a two-component Fermi
gas made up of spin up $\uparrow$ and spin down $\downarrow$ (nondipolar)
fermions of mass $m_{f}\,$\ with population balance [Fig.
\ref{Fig:mixtureInLattice}(b)] in square optical lattices. \ An external field
is introduced to align all dipoles along its direction $\mathbf{\hat{e}}$, a
unit vector specified by elevation and azimuthal angles, $\left(  \theta
_{d},\phi_{d}\right)  $, which are defined with respect to the 2D lattice
plane as illustrated in Fig. \ref{Fig:mixtureInLattice}. \ The system is
assumed to operate in the quasi-2D regime where atoms still experience 3D
interactions which are divided into a short- and long-range part. \ The
short-range part is made up of the s-wave scatterings characterized with the
strength $U_{bb}^{3D}=4\pi\hbar^{2}a_{bb}/m_{b}$, between two bosons,
$U_{bf}^{3D}=4\pi\hbar^{2}a_{bf}/\left[  m_{bf}=2m_{b}m_{f}/\left(
m_{b}+m_{f}\right)  \right]  $, between a boson and a fermion, and
$U_{ff}^{3D}=4\pi\hbar^{2}a_{ff}/m_{F}$, between two fermions of opposite
spin, where $a_{\alpha\beta}$ are the relevant s-wave scattering lengths. (The
s-wave interaction between two fermions of identical spins is prohibited by
the Pauli exclusion principle.) \ The long-range part is the dipole-dipole
interaction between bosons, which takes the form $U_{dd}^{3D}(
\mathbf{r})  =d^{2}[  1-3(  \mathbf{\hat{e}}\cdot
\mathbf{\hat{r}})  ^{2}]  /\left\vert \mathbf{r}\right\vert ^{3}$
in position $\mathbf{r}$ space, with $d$ the induced dipole moment.

The quasi-2D optical lattice is created by three pairs of laser fields of
wavelength $\lambda$ propagating along three orthogonal directions
\cite{petsas94PhysRevA.50.5173,guidoni97PhysRevLett.79.3363} where for
simplicity we have assumed that the laser fields for bosons have the same
wavelength as those for fermions. \ This arrangement creates a trap potential,
$V_{\alpha}\left(  \mathbf{r}\right)  =V_{1\alpha}\left(  z\right)
+V_{2\alpha}\left(  \mathbf{r}_{\perp}\right)  $, for bosons ($\alpha=b$) and
fermions\ ($\alpha=f$), where%
\begin{align}
V_{1\alpha}\left(  z\right)   &  =V_{\alpha z}\sin^{2}\left(  \pi z/a\right)
,\label{V1alpha}\\
V_{2\alpha}\left(  \mathbf{r}_{\perp}\right)   &  =V_{\alpha\perp}\left[
\sin^{2}\left(  \pi x/a\right)  +\sin^{2}\left(  \pi y/a\right)  \right]  .
\label{V2alpha}%
\end{align}
$V_{1\alpha}\left(  z\right)  $ in Eq. (\ref{V1alpha}), with $V_{\alpha z}$
being much stronger than $V_{\alpha\perp}$, is to provide atoms with a tight
confinement along the axial dimension $z$ while $V_{2\alpha}\left(
\mathbf{r}_{\perp}\right)  $ in Eq. (\ref{V2alpha}) is to provide atoms with a
qausi-2D square lattice of lattice constant $a=\lambda/2$ in the radial or
transverse dimensions $\mathbf{r}_{\perp}\mathbf{=}\left(  x,y\right)  $.
\ The parameters are chosen in such a way that $V_{\alpha\perp}$, while much
weaker than $V_{\alpha z}$, is sufficiently high compared with the photon
recoil energy, $E_{\alpha}=2\hbar^{2}\pi^{2}/\lambda^{2}m_{\alpha}$, and all
other energy scales are sufficiently small compared with the energy band gap
(in the order of $V_{\alpha\perp}$). \ As such, atoms reside essentially in
the lowest Bloch band and the tight binding remains a fairly good
approximation. \ In our calculation, we assume that the single-particle
Wannier functions located at site $i$ are in a separable form, $w_{\alpha
}\left(  z\right)  w_{\alpha}\left(  \mathbf{r}_{\perp}\right)  $, where
\begin{align}
w_{\alpha}\left(  z\right)   &  =\exp\left(  -z^{2}/2d_{\alpha z}^{2}\right)
/\sqrt{\sqrt{\pi}d_{\alpha z}},\\
w_{\alpha}\left(  \mathbf{r}_{\perp}\right)   &  =\exp\left(  -\mathbf{r}%
_{\perp}^{2}2d_{\alpha\perp}^{2}\right)  /\sqrt{\pi}d_{\alpha\perp},
\end{align}
are the ground state wavefunctions of the harmonic oscillators, with
$d_{\alpha z,\perp}=\sqrt{\hbar/m_{\alpha}\omega_{\alpha z,\perp}}$ the
harmonic oscillator lengths and $\omega_{\alpha z,\perp}=2\sqrt{E_{\alpha
}V_{\alpha z,\perp}}/\hbar$ the harmonic oscillator frequencies. \ \ The
Bose-Fermi mixture under these approximations is described by the single-band
Bose-Fermi Hamiltonian,%
\begin{align}
\hat{H}  &  =-\sum_{\left\langle ij\right\rangle }\left[  t_{b}\hat{b}%
_{i}^{\dag}\hat{b}_{j}+t_{f}\left(  \hat{a}_{i,\uparrow}^{\dag}\hat
{a}_{j,\uparrow}+\hat{a}_{i,\downarrow}^{\dag}\hat{a}_{j,\downarrow}\right)
-\frac{U_{ij}}{2}\hat{n}_{i,b}\hat{n}_{j,b}\right] \nonumber\\
&  +\sum_{i}\left[  \frac{U_{b}}{2}\hat{n}_{i,b}\left(  \hat{n}_{i,b}%
-1\right)  +U_{bf}\hat{n}_{i,b}\hat{n}_{i,f}+U_{ff}\hat{n}_{i,\uparrow}\hat
{n}_{i,\downarrow}\right] \nonumber\\
&  -\sum_{i}\left[  \mu_{b}\hat{b}_{i}^{\dag}\hat{b}_{i}+\mu_{f}\left(
\hat{a}_{i,\uparrow}^{\dag}\hat{a}_{i,\uparrow}+\hat{a}_{i,\downarrow}^{\dag
}\hat{a}_{i,\downarrow}\right)  \right]  , \label{Hamiltonian}%
\end{align}
where $\left\langle ij\right\rangle $ stands for the summation over nearest
neighbors, $\hat{b}_{i}$ and $\hat{a}_{i,\sigma}$ are the field operators for
bosons and fermions of spin $\sigma\left(  =\uparrow,\downarrow\right)  $ at
site $i$, respectively, $\ \hat{n}_{i,b}=\hat{b}_{i}^{\dag}\hat{b}_{i}$ and
$\hat{n}_{i,\sigma}=\hat{a}_{i,\sigma}^{\dag}\hat{a}_{i,\sigma}$ $\left(
\hat{n}_{i,f}=\hat{n}_{i,\uparrow}+\hat{n}_{i,\downarrow}\right)  $ are the
corresponding particle number operators, and $\mu_{b}$ and $\mu_{f}$ are the
chemical potentials for bosons and fermions, respectively. \ The Hamiltonian
(\ref{Hamiltonian}) has been expressed in terms of various effective 2D
coefficients whose explicit expressions we now present. \ The kinetic energy
part of the Hamiltonian is described by the terms proportional to the hopping
amplitudes between nearest neighbors
\cite{albus03PhysRevA.68.023606,buchler04PhysRevA.69.063603,titvinidze09PhysRevB.79.144506}
\begin{equation}
t_{\alpha}\simeq\frac{4}{\sqrt{\pi}}E_{\alpha}\left(  \frac{V_{\alpha\perp}%
}{E_{\alpha}}\right)  ^{3/4}\exp\left(  -2\sqrt{\frac{V_{\alpha\perp}%
}{E_{\alpha}}}\right)  .
\end{equation}
The on-site interactions consist of
\begin{align}
U_{ff}  &  =U_{ff}^{3D}\left(  \sqrt{2\pi}d_{fz}\right)  ^{-1}\left(  2\pi
d_{f\perp}^{2}\right)  ^{-1},\\
U_{bf}  &  =U_{bf}^{3D}\left(  \sqrt{\pi\left(  d_{bz}^{2}+d_{fz}^{2}\right)
}\right)  ^{-1}\left[  \pi\left(  d_{b\perp}^{2}+d_{f\perp}^{2}\right)
\right]  ^{-1},\\
U_{bb}  &  =U_{bb}^{3D}\left(  \sqrt{2\pi}d_{bz}\right)  ^{-1}\left(  2\pi
d_{b\perp}^{2}\right)  ^{-1},
\end{align}
which are 2D analogs of 3D short-range s-wave interactions, $U_{\alpha\beta
}^{3D}$, introduced in the beginning of the section. \ As to the long-range
dipole-dipole interaction, it contributes both to the onsite interaction and
the off-site interaction. \ \ The former is given by
\cite{yi00PhysRevA.61.041604}
\begin{equation}
U_{dd}=2U_{bb}\times\left[  \chi_{dd}\equiv\varepsilon_{dd}\left(  1-\frac
{3}{2}\sin^{2}\theta_{d}\right)  \zeta\right]  , \label{chi_dd}%
\end{equation}
where $\varepsilon_{dd}=4\pi d^{2}/\left(  3U_{bb}^{3D}\right)  $
measures\ the dipolar interaction relative to the s-wave boson-boson
interaction, and
\begin{equation}
\zeta=1+\frac{3}{2s^{2}}\left(  1-\frac{s^{2}+1}{s}\tan^{-1}s\right)  ,
\label{beta}%
\end{equation}
is a unitless factor fixed purely by the geometry of the trapping potential,
where $s=\sqrt{\sqrt{V_{bz}/V_{b\perp}}-1}$. \ The first term in the second
line of Eq. (\ref{Hamiltonian}) contains
\begin{equation}
U_{b}=U_{bb}+U_{dd}=U_{bb}\left(  1+2\chi_{dd}\right)  ,
\end{equation}
which represents the total onsite interaction for bosons. \ Here, the use of
the single-subscripted $b$ on $U_{b}$ is to stress that $U_{b}$ has
contributions not only from the short-range s-wave interaction $U_{bb}$ but
also from the on-site dipole-dipole interaction $U_{dd}$. \ The off-site
dipole-dipole interaction, owing to their rapid decay with distance ($1/r^{3}%
$), is assumed to be limited to the nearest-neighbor (NN) dipole-dipole
interaction \cite{lin10PhysRevB.81.045115}, well approximated by the formula
\cite{menotti07PhysRevLett.98.235301} $U_{ij}=d^{2}\left(  1-3\cos^{2}%
\theta_{ij}\right)  /a^{3}$, where $\theta_{ij}$ is the angle between the
dipole and the displacement from site $i$ to nearest site $j$.

To study the superfluid pairings, we move to the lattice momentum space (with
a total of $N_{L}$ sites) in which $\hat{b}_{\mathbf{k}}$ and $\hat
{a}_{\mathbf{k},\sigma}$ are field operators and $\hat{\rho}_{\mathbf{k}%
,b}=\sum_{\mathbf{q}}\hat{b}_{\mathbf{q}+\mathbf{k}}^{\dag}\hat{b}%
_{\mathbf{q}}$ and $\hat{\rho}_{\mathbf{k},\sigma}=\sum_{\mathbf{q}}\hat
{a}_{\mathbf{q}+\mathbf{k},\sigma}^{\dag}\hat{a}_{\mathbf{q},\sigma}$ $\left(
\hat{\rho}_{\mathbf{k},f}=\hat{\rho}_{\mathbf{k},\uparrow}+\hat{\rho
}_{\mathbf{k},\downarrow}\right)  $ are the corresponding particle number
operators, and transform the Hamiltonian in Eq. (\ref{Hamiltonian}) into
\begin{align}
\hat{H}  &  =\sum_{\mathbf{k}}\left(  \epsilon_{\mathbf{k},b}-\mu_{b}\right)
\hat{b}_{\mathbf{k}}^{\dag}\hat{b}_{\mathbf{k}}\nonumber\\
&  +\frac{1}{2N_{L}}\sum_{\mathbf{k}}\left[  U_{b}+2U\left(  \mathbf{k}%
\right)  \right]  \hat{\rho}_{\mathbf{k},b}\hat{\rho}_{-\mathbf{k}%
,b}\nonumber\\
&  +\sum_{\mathbf{k\sigma}}\left(  \epsilon_{\mathbf{k},f}-\mu_{f}\right)
\hat{a}_{\mathbf{k,}\sigma}^{\dag}\hat{a}_{\mathbf{k},\sigma}\nonumber\\
&  +\frac{1}{N_{L}}\sum_{\mathbf{k}}\left(  U_{bf}\hat{\rho}_{\mathbf{k}%
,b}\hat{\rho}_{-\mathbf{k},f}+U_{ff}\hat{\rho}_{\mathbf{k},\uparrow}\hat{\rho
}_{-\mathbf{k},\downarrow}\right)  , \label{H1}%
\end{align}
where the sum over the momentum $\mathbf{k}$ (or $\hbar\mathbf{k}$ to be
precise) is limited to the first Brillouin zone,
\begin{equation}
\epsilon_{\mathbf{k},\alpha}=-2t_{\alpha}\left(  \cos k_{x}+\cos k_{y}\right)
\label{single-particle dispersion}%
\end{equation}
is the single-particle dispersion and
\begin{equation}
U\left(  \mathbf{k}\right)  =U_{x}\cos k_{x}+U_{y}\cos k_{y} \label{U(k)}%
\end{equation}
is the NN dipole-dipole interaction in momentum space, with $U_{x}$ and
$U_{y}$ being given by
\begin{align}
U_{x}  &  =\frac{d^{2}}{a^{3}}\left[  1-3\sin^{2}\theta_{d}\cos^{2}\phi
_{d}\right]  ,\\
U_{y}  &  =\frac{d^{2}}{a^{3}}\left[  1-3\sin^{2}\theta_{d}\sin^{2}\phi
_{d}\text{ }\right]  .
\end{align}
We have used the unit convention in which $k_{x}$ and $k_{y}$ are scaled to
the inverse lattice constant $1/a$.

\section{Phonons and Effective Fermi Hamiltonian}

\begin{figure}
[ptb]
\begin{center}
\includegraphics[
width=2.5in
]%
{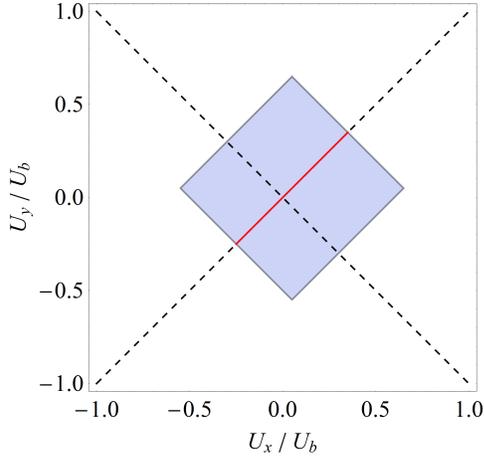}%
\caption{(Color online) The shaded region in the ($U_{x}$,$U_{y}$) plane for
\thinspace$t_{b}/n_{b}U_{b}=1$ supports a stable homogeneous dipolar BEC. For
a system where the NN interaction is isotropic, such a uniform superfluid
exists only along the solid red diagonal line.}%
\label{Fig:regionOfDipolarBEC}%
\end{center}
\end{figure}

An ultracold Bose gas can be prepared in ground states of rather different
natures, e.g., superfluid and Mott insulator phases, depending on the filling
factor and strength of the two-body particle interaction relative to the
hopping energy. \ A number of papers studied Bose-Fermi mixtures in which the
Bose gas operates in the parameter regime close to the Mott
insulator-superfluid transition with a relatively low boson filling factor
(see, e.g.,
references\ \cite{titvinidze08PhysRevLett.100.100401,sinha09PhysRevB.79.115124}%
). \ A dipolar Bose gas is more complex than its nondipolar counterpart and
thus may enter additional phases. \ In our model here, owing to the dipoles
being directed away from the axial direction, \ the dipole-dipole interaction
inherits both side-by-side repulsion and head-to-tail attraction. \ This,
along with the ability to tune the ratios $n_{b}U_{b}/t_{b}$ and $U_{x}/U_{y}%
$, makes not only the superfluid and Mott insulator but also supersolid phases
accessible to the present 2D dipolar system. \ We focus on the Bose-Fermi
mixture in which the Bose gas operates in the superfluid region where the
dipolar superfluid phase is characterized with a uniform order parameter
$\langle \hat{b}_{j}\rangle =\sqrt{n_{b}}$, a chemical potential
$\mu_{b}=-4t_{b}+n_{b}(  U_{b}+2U(  0)  )  $,
and a Bogoliubov excitation spectrum
\begin{equation}
E_{\mathbf{k},b}=\sqrt{\left(  \epsilon_{\mathbf{k},b}+4t_{b}\right)  \left(
\epsilon_{\mathbf{k},b}+4t_{b}+2n_{b}\left[  U_{b}+2U\left(  \mathbf{k}%
\right)  \right]  \right)  }, \label{Ekb}%
\end{equation}
where $U\left(  \mathbf{k}\right)  $ has been defined in Eq. (\ref{U(k)}).
\ This superfluid phase exists in the parameter space where $E_{\mathbf{k},b}$
is real; when $E_{\mathbf{k},b}$ changes from real to imaginary, the dipolar
gas may undergo a dynamical instability towards other competing phases.
\ \ The parameter region that supports a stable homogeneous superfluid in the
$\left(  U_{x},U_{y}\right)  $ space is marked in Fig.
\ref{Fig:regionOfDipolarBEC} as a square bordered by the four lines
$2U_{x}+2U_{y}+U_{b}=0,$ $2U_{x}+2U_{y}-U_{b}-4t_{b}/n_{b}=0,$ $2U_{x}%
-2U_{y}-U_{b}-2t_{b}/n_{b}=0,$ and $2U_{y}-2U_{x}-U_{b}-2t_{b}/n_{b}=0$.
Inside the square, the excitation spectra in the long wavelength limit, at the
mode of the checkerboard supersolid phase $\mathbf{k=}\left(  \pi
/a,\pi/a\right)  $, and at the modes of the striped supersolid phase
$\mathbf{k}=\left(  \pi/a,0\right)  $ and $\left(  0,\pi/a\right)  $, are all
real \cite{danshita09PhysRevLett.103.225301}.%
\ 

As is well-known,\ in a Bose-Fermi mixture, phonons of the condensate can
induce an effective attraction between two fermions
\cite{bijlsma00PhysRevA.61.053601}. \ As a result, in addition to interacting
through the s-wave interaction (between fermions of opposite spins) with strength $U_{ff}$, two fermions can interact indirectly by exchanging virtual
phonons. \ Exchanging virtual phonons in this fashion induces between two
fermions an effective spin-independent attractive interaction, which, in
momentum space, is given by
\begin{equation}
U_{ind}\left(  \mathbf{q}\right)  =-\frac{U_{0}\left(  \equiv U_{bf}%
^{2}/U_{bb}\right)  }{1+2\chi_{dd}+\xi^{2}\left[  4-2D\left(  \mathbf{k}%
\right)  \right]  }, \label{Uind}%
\end{equation}
where $D\left(  \mathbf{k}\right)  =\sum_{\alpha=x,y}\left(  1-2n_{b}%
U_{\alpha}/t_{b}\right)  \cos k_{\alpha}$ and $\ \xi=\sqrt{t_{b}/2n_{b}U_{bb}%
}$ is the healing length (in the unit of lattice constant $a$) of the bosons
interacting only under the short-range s-wave interaction. In
arriving at Eq. (\ref{Uind}), we have ignored the retardation effect. \ This
implies that Eq. (\ref{Uind}) holds when the characteristic energy scale of
fermions is small in comparison to that of phonons, which translates into the
fast phonon limit. \ In this limit, the phonon velocity is much larger than
the Fermi velocity so that the phonon degrees of freedom can be adiabatically
eliminated, leading to Eq. (\ref{Uind}). However, the same formula is often
used in studies where the fast phonon limit is not quite fulfilled. The
rational behind this is that the retardation effect affects the
pre-exponential factor from the Fermi energy scale to some characteristic
bosonic frequency scale \cite{schrieffer64Book,wang05PhysRevA.72.051604}. Thus, \ we expect it to
be true for our dipolar Bose-Fermi mixture as well---including the retardation
effect will not affect the (qualitative) predictive power of the formulas we
aim to derive in this work. 

The effective Hamiltonian for fermions after integrating out the phonon
degrees of freedom becomes, apart from a c-number,
\begin{align}
\hat{H}_{F}  &  =\sum_{\mathbf{k},\sigma}\left(  \epsilon_{\mathbf{k},f}%
-\mu_{f}+n_{b}U_{bf}\right)  \hat{a}_{\mathbf{k},\sigma}^{\dag}\hat
{a}_{\mathbf{k},\sigma}\nonumber\\
&  +\frac{1}{2N_{L}}\sum_{\mathbf{q,\sigma,\sigma}^{\prime}}U_{\sigma
,\sigma^{\prime}}\left(  \mathbf{q}\right)  \hat{\rho}_{\mathbf{q},\sigma}%
\hat{\rho}_{-\mathbf{q},\sigma^{\prime}}, \label{HF_Eff}%
\end{align}
where
\begin{equation}
U_{\sigma,\sigma^{\prime}}\left(  \mathbf{q}\right)  =U_{ff}\delta
_{\sigma^{\prime},-\sigma}+U_{ind}\left(  \mathbf{q}\right)  
\end{equation}
is the total effective Fermi-Fermi interaction.  We shall make two important comments concerning our assumptions in arriving at this effective Hamiltonian.

First, we have taken into account only the effect of bosons on
fermions in the form of the phonon-mediated interaction. In the presence of
Bose-Fermi interaction, fermions will always disturb and therefore modify the
Bose gas. This will result in a correction to the Bose gas proportional to the
Bose-Fermi interaction. This correction will, in turn, affect the Fermi gas,
which, in terms of the Bose-Fermi interaction, is a higher order
effect than the effect of the undisturbed Bose gas on the Fermi gas.
Therefore, one can ignore the former effect in comparison to the latter one in
the limit of weak Bose-Fermi interaction.

Second, we have neglected the fermion-induced interaction, in which a
fermion interacts with a second fermion via polarization of the fermionic
background by the second fermion. Gor'kov and Melik-Barkhudarov
\cite{gorkovSov61PhysJETP:13.1018} found this interaction to be responsible
for an order unity correction to the critical temperature of the superfluid
with s-wave symmetry in Fermi gases with bare attractive interactions (also
see \cite{heiselberg00PhysRevLett.85.2418}).  The same induced-interaction was
proposed by\ Kohn and Luttinger \cite{kohn65PhysRevLett.15.524} as a pairing mechanism
(Kohn-Luttinger mechanism) for realizing superfluids with high-partial wave
symmetries in Fermi gases with bare repulsive interactions.  As such, this
induced interaction plays an important role in the exploration of pairing
physics in pure Fermi gases.  In a Bose-Fermi mixture, however, not only
background fermions but also phonons of the BEC can mediate an interaction
between fermions.  The former interaction can be ignored in situations where it is much weaker than the latter interaction
\cite{efremov02PhysRevB.65.134519}, which will be our case here.  Additionally, we may ignore the fermion interaction since the Kohn-Luttinger effect in 2D systems with rotational invariance is exponentially suppressed compared with its 3D counterpart \cite{galitski03PhysRevB.67.144520}, and our model takes place in the setting of a 2D lattice.  At any rate, our results in this
paper, like those in many others (see, for example,
\cite{viverit02PhysRevA.66.023605, wang05PhysRevA.72.051604, bukov14PhysRevB.89.094502}) which also ignore the Kohn-Luttinger
mechanism, hold in the parameter space where the phonon-induced interaction
dominates.

\section{Superfluid States}

At temperatures below the Fermi temperature, fermions tend to form\ the normal
ground state (Fermi sea), in which states below the Fermi energy are all
occupied with each state accommodating one particle. \ As in crystal solids,
the phonon-mediated interaction in our model, being attractive, may serve as a
catalyst, causing the normal state to undergo dynamical instability towards
the BCS state, in which fermions are divided into highly correlated pairs with
opposite momenta. \ This normal-to-superfluid phase-transition is accompanied
with a lowering of symmetry and is therefore of second-order in nature. \ It
is characterized by a symmetry-breaking order parameter, known as the gap,
that represents the extent of macroscopic phase coherence. \ 

To study the BCS state of a fermionic system described by the effective
Hamiltonian (\ref{HF_Eff}), we use the self-consistent Hartree-Fock-Bogoliubov
mean-field approach \cite{deGennes89Book}, in which we associate the BCS
paring with the gap parameter $\Delta_{\sigma,\sigma^{\prime}}\left(
\mathbf{k}\right)  =\sum_{\mathbf{q}}U_{\sigma,\sigma^{\prime}}\left(
\mathbf{k}-\mathbf{q}\right)  \left\langle \hat{a}_{-\mathbf{q},\sigma}\hat
{a}_{\mathbf{q},\sigma^{\prime}}\right\rangle /N_{L}$, direct pairing with the
Hartree potential $\Sigma^{^{\prime}}=\sum_{\sigma^{\prime}}U_{\sigma
,\sigma^{\prime}}\left(  0\right)  n_{\sigma^{\prime}}$, and exchange paring
with the Fock potential $\Sigma\left(  \mathbf{k}\right)  =-\sum_{\mathbf{q}%
}U_{ind}\left(  \mathbf{k}-\mathbf{q}\right)  \left\langle \hat{a}%
_{\mathbf{q},\sigma}^{\dag}\hat{a}_{\mathbf{q},\sigma}\right\rangle /N_{L}$.
\ A further application of the Bogoliubov diagonalization procedure leads to
the gap equation%
\begin{equation}
\Delta_{\sigma,\sigma^{\prime}}\left(  \mathbf{k}\right)  =-\sum_{\mathbf{q}%
}\frac{U_{\sigma,\sigma^{\prime}}\left(  \mathbf{k}-\mathbf{q}\right)  }%
{N_{L}}\frac{\tanh\frac{E_{\mathbf{q}}}{2T}}{2E_{\mathbf{q}}}\Delta
_{\sigma,\sigma^{\prime}}\left(  \mathbf{q}\right)  , \label{gap 1}%
\end{equation}
and the particle number equation%
\begin{equation}
n_{f}=\frac{1}{N_{L}}\sum_{\mathbf{q}}\left(  1-\frac{\xi_{\mathbf{q}}%
^{\prime}}{E_{\mathbf{q}}}\tanh\frac{E_{\mathbf{q}}}{2T}\right)  ,
\end{equation}
where $\xi_{\mathbf{k}}^{\prime}=\left(  \xi_{\mathbf{k}}\equiv\epsilon
_{\mathbf{k},f}-\mu_{f}\right)  +\Sigma\left(  \mathbf{k}\right)  $, and
$E_{\mathbf{k}}=\sqrt{\xi_{\mathbf{k}}^{\prime2}+\Delta_{\sigma,\sigma
^{\prime}}^{2}\left(  \mathbf{k}\right)  \text{ }}$ is the fermionic
quasi-particle spectrum. \ In Eq. (\ref{gap 1}), we have adjusted the chemical
potential according to $\mu_{f}\rightarrow\mu_{f}-\left[  U_{bf}n_{b}%
+U_{ind}\left(  0\right)  n_{f}+U_{ff}n_{f}/2\right]  $, where $n_{f}%
=n_{\uparrow}+n_{\downarrow}$ is the total density of fermions, which is also
known as the fermion filling factor. \ Note that in this paper, we absorb the
Boltzmann constant $k_{B}$ into the temperature $T$ so that $T$ has the units
of energy. \ 

The gap parameter $\Delta_{\sigma,\sigma^{\prime}}\left(  \mathbf{q}\right)  $
in Eq. (\ref{gap 1}), according to the Landau theory of second order phase
transitions \cite{landau79Book}, must transform according to one of the
irreducible representations of the symmetry group of the high-temperature
disordered phase. \ The symmetry group to which the effective Hamiltonian
$\hat{H}_{F}$ in Eq. (\ref{HF_Eff}) belongs is dictated, in our model where
the hopping is assumed to be isotropic, by the symmetry underlying the
Fermi-Fermi interaction. \ In this work, we explore the superfluid phases of
the effective Hamiltonian $\hat{H}_{F}$ [Eq. (\ref{HF_Eff})] that preserves
the symmetry of a square lattice. \ Hence, we limit our study to systems that
operate, within the (shaded) region in the $\left(  U_{x},U_{y}\right)  $
space of Fig. \ref{Fig:regionOfDipolarBEC}, along the +45$^{%
{{}^\circ}%
}$(red) line, where $U_{x}=U_{y}$ and the NN interaction is isotropic. \ This
can be achieved by fixing the azimuthal angle of the dipole to $\phi_{d}=45^{%
{{}^\circ}%
}$ while changing either the elevation angle $\theta_{d}$ or the induced
dipole moment $d$.  For such systems, the phonon-mediated Fermi-Fermi
interaction, Eq. (\ref{Uind}), simplifies to%
\begin{equation}
U_{ind}\left(  \mathbf{k}\right)  =-\frac{U_{0}}{1+2\chi_{dd}+\xi^{2}\left[
4-b\gamma_{\mathbf{k}}\right]  }, \label{Uind 1}%
\end{equation}
where%
\begin{equation}
b=1-\frac{1}{2}\frac{\chi_{dd}}{\xi^{2}}\times\left[  G\equiv\frac{3}{2}%
\frac{U_{bb}^{3D}}{U_{bb}}\frac{1}{\pi a^{3}\zeta}\right]  , \label{b}%
\end{equation}
and
\begin{equation}
\gamma_{\mathbf{k}}=2\left(  \cos k_{x}+\cos k_{y}\right)  . \label{gamma k}%
\end{equation}
The table in Fig.\ 3 lists the irreducible representations of the point
group $D_{4}$ along with the simplest possible base functions (for a
square lattice) that transform according to the corresponding representations
\cite{kohn65PhysRevLett.15.524}.  We stress that for each irreducible representation, there
corresponds infinitely many other base functions. However, near the Fermi
surface, the quasiparticle energy is determined by the gap parameter (which is
a function of $k$).  It is thus energetically favorable for the gap
parameter to have as few nodes as possible on the Fermi surface, suggesting
that the pairing phases that will most likely materialize are those with the
simplest possible base functions  \cite{schaffer14JPhysCondensMatter.26.423201,gilmore15}. This may explain
why well-known works (see, for example, review articles \cite{scalapino95PhysRep.250.329} and \cite{tsuei00RevModPhys.72.969}) use
the simplest base functions listed in Fig. 3 to classify superfluid order
parameters.

As can be seen, the function $\gamma
_{\mathbf{k}}$ in Eq. (\ref{Uind 1}) transforms according to the total
symmetric representation, $A_{1}$. \ Hence, as anticipated, the interaction in
Eq. (\ref{Uind 1}) exhibits the same symmetry as that of the underlying
optical lattice which, for a square lattice, is described by the point group
$D_{4}$.%
\begin{figure}
[ptb]
\begin{center}
\includegraphics[
width=3.2in
]%
{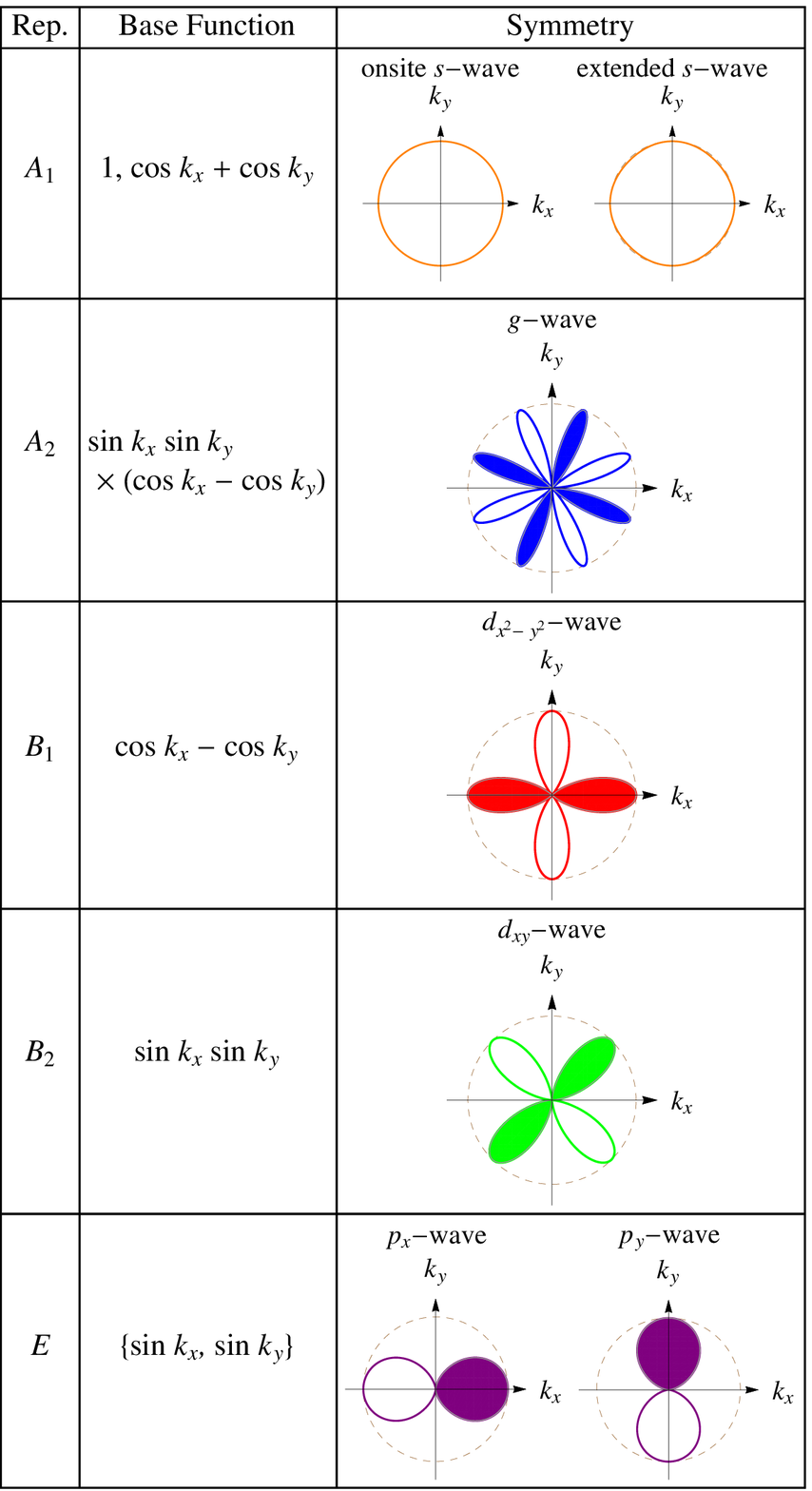}%
\caption{(Color online) The irreducible representations of symmetry group
$D_{4}$ along with their base functions. }%
\label{Fig:D4}%
\end{center}
\end{figure}

In the spirit of the Landau theory of second-order phase transitions, we now
expand the gap parameter in the space spanned by the base functions of the
point group $D_{4}$ in Fig. \ref{Fig:D4},
\begin{align}
\Delta_{\sigma,\sigma^{\prime}}\left(  \mathbf{k}\right)   &  =\delta
_{\sigma,-\sigma^{\prime}}\left(  \Delta_{s0}+\Delta_{s1}\gamma_{\mathbf{k}%
}+\Delta_{d_{x^{2}-y^{2}}}\eta_{\mathbf{k}}\right) \nonumber\\
&  +\delta_{\sigma,-\sigma^{\prime}}\left(  \Delta_{d_{xy}}\beta_{\mathbf{k}%
}+\Delta_{g}\beta_{\mathbf{k}}\eta_{\mathbf{k}}\right) \nonumber\\
&  +\delta_{\sigma,\sigma^{\prime}}\left(  \Delta_{px}\sin k_{x}+\Delta
_{py}\sin k_{y}\right)  , \label{decomposition}%
\end{align}
where $\gamma_{\mathbf{k}}$ has been defined previously in Eq.
(\ref{gamma k}) and%
\begin{equation}
\eta_{\mathbf{k}}=2\left(  \cos k_{x}-\cos k_{y}\right)  ,
\quad
\beta_{\mathbf{k}%
}=\sin k_{x}\sin k_{y}.
\end{equation}
The gap parameter behaves very much like the wave function for two identical
fermions. \ It is made up of two parts: a spin\ part and a spatial part. \ The
first two lines in Eq. (\ref{decomposition}) are devoted to the superfluids in
the spin singlet sector where the spin part is in the singlet state
antisymmetric with respect to spin exchange and the spatial part must then
transform according to the four parity-odd 1D representations $A_{1,2}$ and
$B_{1,2}$ in order to meet the Fermi statistics. \ In contrast, the last line
in Eq. (\ref{decomposition}) is devoted to the superfluids in the spin triplet
sector where the spin part is in the triplet state symmetric with respect to
spin exchange and the spatial part must then transform according to the
parity-even 2D representation $E$. \ Accordingly, $\Delta_{s0}$, $\Delta_{s1}%
$, $\Delta_{d_{x^{2}-y^{2}}}$, $\Delta_{d_{xy}}$ and $\Delta_{g}$ in Eq.
(\ref{decomposition}) are the gap parameters for the superfluids in the spin
singlet sector with on-site $s$-, extended $s$-, $d_{x^{2}-y^{2}}$-, $d_{xy}%
$-, and $g$-wave symmetries, respectively, while $\Delta_{px}$ and
$\Delta_{py}$ in Eq. (\ref{decomposition}) are the gap parameters for the
superfluids in the spin triplet sector with $p$-wave symmetries.

The gap parameter is vanishingly small near the critical temperature. \ In
order to determine the critical temperature for each superfluid state, we
insert the ansatz (\ref{decomposition}) into the gap equation (\ref{gap 1})
linearized around $\Delta_{\sigma,\sigma^{\prime}}\left(  \mathbf{q}\right)
=0$ and replace $U_{\sigma,\sigma^{\prime}}\left(  \mathbf{k},\mathbf{q}%
\right)  $ either with the singlet potential,
\begin{equation}
U_{s}\left(  \mathbf{k},\mathbf{q}\right)  =U_{ff}+\frac{U_{ind}\left(
\mathbf{k}-\mathbf{q}\right)  +U_{ind}\left(  \mathbf{k+q}\right)  }{2},
\label{Us}%
\end{equation}
or with the triplet potential,
\begin{equation}
U_{t}\left(  \mathbf{k},\mathbf{q}\right)  =\frac{U_{ind}\left(
\mathbf{k}-\mathbf{q}\right)  -U_{ind}\left(  \mathbf{k+q}\right)  }{2},
\label{Ut}%
\end{equation}
depending on whether the gap equation falls into the singlet or triplet sector
of superfluid pairings. \ \ This procedure is equivalent to finding, in the
space spanned by the base functions of the irreducible representations of the
point group $D_{4}$, the matrix representation of the linearized gap equation,
which, due to the inherent symmetry of our model, is expected to be block
diagonal. \ For organizational purposes, we divide the gap equation in the
spin singlet sector into one for the s-wave pairing,
\begin{equation}
\left(
\begin{array}
[c]{cc}%
s_{0}^{0}-1 & s_{0}^{1}\\
s_{1}^{0} & s_{1}^{1}-1
\end{array}
\right)  \left(
\begin{array}
[c]{c}%
\Delta_{s0}\\
\Delta_{s1}%
\end{array}
\right)  =0,
\end{equation}
where
\begin{subequations}
\label{s}%
\begin{align}
s_{0}^{0}  &  =-\sum_{\mathbf{k},\mathbf{q}}U_{s}\left(  \mathbf{k}%
,\mathbf{q}\right)  S_{\mathbf{q}}\left(  T\right)  ,\\
s_{0}^{1}  &  =-\sum_{\mathbf{k},\mathbf{q}}U_{s}\left(  \mathbf{k}%
,\mathbf{q}\right)  S_{\mathbf{q}}\left(  T\right)  \gamma_{\mathbf{q}},\\
s_{1}^{0}  &  =-\sum_{\mathbf{k},\mathbf{q}}U_{s}\left(  \mathbf{k}%
,\mathbf{q}\right)  S_{\mathbf{q}}\left(  T\right)  \frac{\gamma_{\mathbf{k}}%
}{4},\\
s_{1}^{1}  &  =-\sum_{\mathbf{k},\mathbf{q}}U_{s}\left(  \mathbf{k}%
,\mathbf{q}\right)  S_{\mathbf{q}}\left(  T\right)  \frac{\gamma_{\mathbf{k}%
}\gamma_{\mathbf{q}}}{4},
\end{align}
and one for the $d_{x^{2}-y^{2}}$-, $d_{xy}$-, and $g$-wave pairings,%
\end{subequations}
\begin{equation}
\left(
\begin{array}
[c]{ccc}%
d_{0}-1 & 0 & 0\\
0 & d_{1}-1 & 0\\
0 & 0 & g-1
\end{array}
\right)  \left(
\begin{array}
[c]{c}%
\Delta_{d_{x^{2}-y^{2}}}\\
\Delta_{d_{xy}}\\
\Delta_{g}%
\end{array}
\right)  =0,
\end{equation}
where
\begin{subequations}
\label{d}%
\begin{align}
d_{0}  &  =-\sum_{\mathbf{k},\mathbf{q}}U_{s}\left(  \mathbf{k},\mathbf{q}%
\right)  S_{\mathbf{q}}\left(  T\right)  \frac{\eta_{\mathbf{k}}%
\eta_{\mathbf{q}}}{4},\label{d0}\\
d_{1}  &  =-\sum_{\mathbf{k},\mathbf{q}}U_{s}\left(  \mathbf{k},\mathbf{q}%
\right)  S_{\mathbf{q}}\left(  T\right)  4\beta_{\mathbf{k}}\beta_{\mathbf{q}%
},\label{d1}\\
g  &  =-\sum_{\mathbf{k},\mathbf{q}}U_{s}\left(  \mathbf{k},\mathbf{q}\right)
S_{\mathbf{q}}\left(  T\right)  2\beta_{\mathbf{k}}\eta_{\mathbf{k}}%
\beta_{\mathbf{q}}\eta_{\mathbf{q}}, \label{g}%
\end{align}
with $S_{\mathbf{q}}\left(  T\right)  $ being defined as%
\end{subequations}
\begin{equation}
S_{\mathbf{q}}\left(  T\right)  =\tanh\left(  \xi_{\mathbf{q}}^{\prime
}/2T\right)  /2\xi_{\mathbf{q}}^{\prime}. \label{Sq(T)}%
\end{equation}
The gap equation in the spin triplet sector is also diagonalized with the form
given by%
\begin{equation}
\left(
\begin{array}
[c]{cc}%
p_{x}-1 & 0\\
0 & p_{y}-1
\end{array}
\right)  \left(
\begin{array}
[c]{c}%
\Delta_{px}\\
\Delta_{py}%
\end{array}
\right)  =0,
\end{equation}
where, due to symmetry, the coefficients $p_{x}=p_{y}\equiv p$, with $p$ being
given by
\begin{equation}
p=-\sum_{\mathbf{k},\mathbf{q}}U_{t}\left(  \mathbf{k},\mathbf{q}\right)
S_{\mathbf{q}}\left(  T\right)  2\sin k_{x}\sin q_{x}. \label{triplet}%
\end{equation}

The coefficients in Eqs. (\ref{s}), (\ref{d}) and (\ref{triplet}) are nothing
but the matrix elements [of the linearized gap equation (\ref{gap 1}), or, to
be more precise, the right-hand side of Eq. (\ref{gap 1})] between various
base functions of the irreducible representations of $D_{4}$. \ \ A major
complication in evaluating these coefficients is that they are functions of
the Fock potential $\Sigma\left(  \mathbf{k}\right)  $ and chemical potential
$\mu_{f}$, which, in principle, are themselves unknowns to be determined
self-consistently. \ We take two measures to reduce this complexity. \ First,
we estimate the Fock potential on the Fermi surface and absorb it into the
chemical potential. \ Second, we compute the chemical potential from the
number equation for the $T=0$ normal Fermi gas,%
\begin{equation}
n_{f}=\frac{4}{\pi^{2}}\int_{-1}^{\bar{\mu}_{f}}K\left(  1-x^{2}\right)
dx,\label{nf}%
\end{equation}
where $\bar{\mu}_{f}$ $=\mu_{f}/4t_{f}$ is the scaled chemical potential and
$K\left(  x\right)  $ is the complete elliptic integral of the first kind.
\ (All elliptical integrals in this paper follow the convention of Ref.
\cite{abramowitz64}.) \ It is to be stressed that only in the weak
interacting limit where the characteristic interaction energy is much weaker
than the Fermi energy (which is roughly equal to $4t_{f}$ when near half
filling) do both approximations hold. \ Thus, the expressions we will show
below remain quantitatively accurate only in the weak coupling regime, but,
nevertheless, are expected to help us gain some qualitative insights into
the superfluid pairings under more general conditions. \ \ Note that these
were the same approximations Micnas and Ranninger
\cite{micnas88PhysRevB.37.9410} used when they needed to gain quick insights
into their critical temperatures.

In spite of these assumptions, these matrix elements remain complicated
integrals in 4D momentum space involving functions, $U_{s,t}\left(
\mathbf{k},\mathbf{q}\right)  $, $S_{\mathbf{q}}\left(  T\right)  $, etc.,
that have nontrivial dependences on momenta. \ Nevertheless, as the example in the
Appendix illustrates, we are able to reduce each integral into a product
between a 1D integral and an analytical function. \ The 1D integral (to be
defined shortly) is with respect to $x\equiv\epsilon_{\mathbf{k},f}/4t_{f}$
and involves
\begin{equation}
S_{\pm}\left(  x,\bar{T}\right)  =S\left(  x,\bar{T}\right)  \pm S\left(
-x,\bar{T}\right)  ,
\end{equation}
where
\begin{equation}
S\left(  x,\bar{T}\right)  =\frac{\tanh\left(  \frac{x-\bar{\mu}_{f}}{2\bar
{T}}\right)  }{2\left(  x-\bar{\mu}_{f}\right)  }, \label{S(x,T)}%
\end{equation}
is $S_{\mathbf{q}}\left(  T\right)  $ of Eq. (\ref{Sq(T)}) in the scaled form.
\ The analytical function will be one of the following functions:
\begin{subequations}
\label{I}%
\begin{align}
I_{1}  &  =\frac{1}{\pi^{3}}\text{ }\left[  \frac{\bar{U}_{0}\chi}{b\xi^{2}%
}K\left(  \chi^{2}\right)  -2\pi\bar{U}_{ff}\right]  ,\label{I1}\\
I_{2}  &  =\frac{\bar{U}_{0}\left[  K\left(  \chi^{2}\right)  -\frac{\pi}%
{2}\right]  }{\pi^{3}b\xi^{2}},\label{I2}\\
I_{3}  &  =\frac{8}{3\pi^{3}}\frac{\bar{U}_{0}}{b\xi^{2}}\left[  \left(
\frac{4}{\chi}-2\chi\right)  K\left(  \chi^{2}\right)  -\frac{4}{\chi}E\left(
\chi^{2}\right)  \right]  ,\label{I3}\\
I_{4}  &  =\frac{32}{15\pi^{3}}\frac{\bar{U}_{0}}{b\xi^{2}}\left\{
2\pi-\left(  \frac{4}{\chi}\right)  ^{2}E\left(  \chi^{2}\right)  \right.
\text{ \ }\label{I4}\\
&  \left.  +\left[  \left(  \frac{4}{\chi}\right)  ^{2}-12\right]  K\left(
\chi^{2}\right)  \right\}  ,
\end{align}
where $K\left(  x\right)  $ and $E\left(  x\right)  $ are, respectively,
complete elliptical integrals of the first and second kind, and $\chi$ is a
function defined as
\end{subequations}
\begin{equation}
\chi=\frac{4b\xi^{2}}{1+2\chi_{dd}+4\xi^{2}}.
\end{equation}
Here, the scaled variables, $\bar{U}_{0}=U_{0}/4t_{f}$, $\bar{U}_{ff}%
=U_{ff}/4t_{f}$, $\bar{T}=T/4t_{f}$, are similarly defined as the scaled
chemical potential $\bar{\mu}_{f}$ introduced in Eq. (\ref{nf}).

The conditions for the onset of various superfluid pairings can then be
expressed in terms of various 1D integrals that we now introduce.  The
critical temperature is determined from
\begin{equation}
\det\left\vert
\begin{array}[c]{cc}
s_{0}^{0}\left(  \bar{T}\right)  -1 & s_{0}^{1}\left(  \bar{T}\right) \\
s_{1}^{0}\left(  \bar{T}\right)  & s_{1}^{1}\left(  \bar{T}\right)  -1
\end{array}
\right\vert 
=0, 
\label{s-wave superfluid}%
\end{equation}
for the pairing with the $s$-wave symmetry, where
\begin{allowdisplaybreaks}
\begin{subequations}
\label{s integtrals}
\begin{align}
s_{0}^{0}  &  =I_{1}\int_{0}^{1}K\left(  1-x^{2}\right)  S_{+}\left(
x,\bar{T}\right)  dx,\\
s_{0}^{1}  &  =-I_{1}\int_{0}^{1}4xK\left(  1-x^{2}\right)  S_{-}\left(
x,\bar{T}\right)  dx,\\
s_{1}^{0}  &  =-I_{2}\int_{0}^{1}xK\left(  1-x^{2}\right)  S_{-}\left(
x,\bar{T}\right)  dx,\\
s_{1}^{1}  &  =I_{2}\int_{0}^{1}4x^{2}K\left(  1-x^{2}\right)  S_{+}\left(
x,\bar{T}\right)  dx,
\end{align}
from
\end{subequations}
\end{allowdisplaybreaks}
\begin{equation}
p\left(  \bar{T}\right)  =1,
\end{equation}
for the $p$-wave pairing, where
\begin{equation}
p=I_{2}\int_{0}^{1}4\left[  E\left(  1-x^{2}\right)  -x^{2}K\left(
1-x^{2}\right)  \right]  S_{+}\left(  x,T\right)  dx, \label{p integrals}%
\end{equation}
and finally from%
\begin{equation}
d_{0}\left(  \bar{T}\right)  =1,
\quad d_{1}\left(  \bar{T}\right)  =1,
\quad g\left(
\bar{T}\right)  =1,
\end{equation}
for the pairings with $d_{x^{2}-y^{2}}$-, $d_{xy}$-, and $g$-wave symmetries,
respectively, where
\begin{widetext}%
%

\begin{align}
d_{0}  &  =I_{2}\int_{0}^{1}dxS_{+}\left(  x,\bar{T}\right)  8\left(
1+x\right)  \left\{  K\left[  \left(  \frac{1-x}{1+x}\right)  ^{2}\right]
-E\left[  \left(  \frac{1-x}{1+x}\right)  ^{2}\right]  \right\}  ,\label{dd}\\
d_{1}  &  =I_{3}\int_{0}^{1}dxS_{+}\left(  x,\bar{T}\right)  \left(
1+x\right)  \left\{  \left(  1+x^{2}\right)  E\left[  \left(  \frac{1-x}%
{1+x}\right)  ^{2}\right]  -2xK\left[  \left(  \frac{1-x}{1+x}\right)
^{2}\right]  \right\}  ,\label{ee}\\
g  &  =I_{4}\int_{0}^{1}dxS_{+}\left(  x,\bar{T}\right)  \left\{  \left(
1+14x^{2}+x^{4}\right)  \left(  1+x\right)  E\left[  \left(  \frac{1-x}%
{1+x}\right)  ^{2}\right]  -2x\left(  1+x\right)  \left(  1+6x+x^{2}\right)
K\left[  \left(  \frac{1-x}{1+x}\right)  ^{2}\right]  \right\}  . \label{gg}%
\end{align}%
\end{widetext}%
The reduction of 4D integrals into 1D integrals, which constitute the main
results of the present work, involves an extensive use of algebra containing
elliptical integrals and symmetry considerations. \ In the Appendix, we show
how to simplify the matrix element for the d-wave pairing from Eq.
(\ref{d0})~into Eq. (\ref{dd}), using it as an example to showcase the
techniques that help us simplify the matrix elements for all other pairings. The reasons why we consider these results significant will be
presented in the final section (Sec. VII) where we conclude this work.

\section{CDW and SDW States}

The subject of fermionic superfluids studied in the previous section is
founded on an instability discovered by Cooper --- in the presence of an
attractive interaction, irrespective of its weakness, two fermions on a Fermi
surface find it energetically favorable to form a bound state, thereby causing
a normal Fermi gas to undergo dynamical instability towards the BCS state.
\ This, however, is not the only instability which a normal Fermi gas may
experience. \ In our model, fermions are subject to the density-density
interaction, $N_{L}^{-1}\sum_{\mathbf{k}}U_{bf}\hat{\rho}_{\mathbf{k},b}%
\hat{\rho}_{-\mathbf{k},f}$, due to the s-wave scattering between bosons and
fermions. \ This interaction induces a change in the fermion density, which,
within linear response theory, equals $2U_{bf}\chi\left(  {\mathbf{q}}\right)
\hat{n}_{b}\left(  \mathbf{q}\right)  $ for a two-component Fermi gas, where%

\begin{equation}
\chi\left(  \mathbf{q},T\right)  =\frac{1}{N_{L}}\sum_{\mathbf{k}}%
\frac{f\left(  \xi_{\mathbf{k}}\right)  -f\left(  \xi_{\mathbf{k}+\mathbf{q}%
}\right)  }{\xi_{\mathbf{k}}-\xi_{\mathbf{k}+\mathbf{q}}+i\eta},
\label{Lindhard}%
\end{equation}
is the well-known Lindhard function (in the static limit) with $f\left(
x\right)  =\left[  1+\exp\left(  x/T\right)  \right]  ^{-1}$ the Fermi
distribution function. \ 

For continuous models, no drastic response of this function occurs at any
$\mathbf{q}$ in 3D and 2D, but does in 1D, where it is found to diverge
logarithmically at $q=Q\equiv2k_{F}$. \ \ The Fermi surface of a 1D Fermi gas
consists of two points, one at $+k_{F}$ and the other at $-k_{F}$. \ This
along with the dispersion relation being linear near the Fermi points results
in a perfect nesting, where the two Fermi points can be mapped to each other
via the nesting condition, $\xi_{\mathbf{k+Q}}=-\xi_{\mathbf{k}}$, thereby
giving a divergent contribution to the Lindhard function in Eq.
(\ref{Lindhard}) \cite{gruner94DensityWavesInSolids}. \ 

For discrete (lattice) 2D models, owing to the reduced symmetries (relative to
the continuous 2D models), similar nesting effects can arise. \ In particular,
nesting is known to occur at the nesting vector $\mathbf{Q=}\left(  \pi
,\pi\right)  $ for a half-filled Fermi gas in a square lattice, leading to the
divergence of $\chi\left(  \mathbf{Q},T\rightarrow0\right)  $ at half filling
\cite{buchler03PhysRevLett.91.130404,orth09PhysRevA.80.023624}. \ This can
also be understood from the zero temperature Lindhard susceptibility,
$\chi\left(  \mathbf{Q},T=0\right)  $, as a function of the chemical potential
$\mu_{f}$. \ Replacing $f\left(  \xi_{\mathbf{k}}\right)  $ with
$\Theta\left(  -\xi_{\mathbf{k}}\right)  $ at $T=0$, we can simplify Eq.
(\ref{Lindhard}) into
\[
\chi\left(  \mathbf{Q},T=0\right)  =\frac{1}{2t_{f}\pi^{2}}\int_{-1}^{\mu
_{f}/4t_{f}}\frac{dx}{x}K\left[  1-x^{2}\right]  .
\]
The singularity, $1/x$, due to nesting and the van Hover singularity,
$K\left(  1-x^{2}\right)  \approx\ln\left(  4/\left\vert x\right\vert \right)
$, in the density of states combine to yield a divergent response of the
susceptibility, in the limit of half filling where $\mu_{f}\rightarrow0$,
according to
\[
\chi\left(  \mathbf{Q},T=0\right)  \approx-\frac{1}{4t_{f}\pi^{2}}\left(
\ln\left\vert \frac{16t_{f}}{\mu_{f}}\right\vert \right)  ^{2}.
\]
This divergence implies that at very low temperatures and when operating close
to half filling, a normal Fermi gas is unstable against spatially varying
density perturbations and may evolve spontaneously into phases that may rival
the superfluid phases studied in the previous section. We focus on two such
possibilities, the spin-density wave (SDW) characterized with order parameter
\begin{equation}
\Delta_{SDW}=\frac{1}{2N_{L}}\sum_{\mathbf{k},\sigma}\sigma\left\langle
\hat{c}_{\mathbf{k},\sigma}^{\dag}\hat{c}_{\mathbf{k}+\mathbf{Q},\sigma
}\right\rangle ,
\end{equation}
where the spin density varies in space with a commensurate wavevector
$\mathbf{Q}$ (the case of an incommensurate wavevector is beyond the scope of
this paper), and the charge-density wave (CDW) characterized with order parameter,%

\begin{equation}
\Delta_{CDW}=\frac{1}{2N_{L}}\sum_{\mathbf{k},\sigma}\left\langle \hat
{c}_{\mathbf{k},\sigma}^{\dag}\hat{c}_{\mathbf{k}+\mathbf{Q},\sigma
}\right\rangle ,
\end{equation}
where the charge density modulates in space with a commensurate wavevector
$\mathbf{Q}$.

In the same spirit as the Hartree-Fock Bogoliubov mean-field approach that we
employed previously in the study of superfluid pairings, we arrive at the
mean-field equation for the order parameter $\Delta_{SDW}$
\begin{equation}
1=\frac{U_{ff}}{4N_{L}}\sum_{\mathbf{k}}\frac{\tanh\frac{E_{\mathbf{k},+}}%
{2T}-\tanh\frac{E_{\mathbf{k},-}}{2T}}{\sqrt{\left(  \frac{\xi_{\mathbf{k}%
,\uparrow}^{\prime}-\xi_{\mathbf{k}+\mathbf{Q},\uparrow}^{\prime}}{2}\right)
^{2}+U_{ff}^{2}\left\vert \Delta_{SDW}\right\vert ^{2}}},
\end{equation}
and that for the particle number density%
\begin{equation}
n_{\sigma}=1-\frac{1}{2N_{L}}\sum_{\mathbf{k}}\left(  \tanh\frac
{E_{\mathbf{k},+}}{2T}+\tanh\frac{E_{\mathbf{k},-}}{2T}\right)  ,
\end{equation}
where
\begin{align}
E_{\mathbf{k},\pm}  &  =\frac{\xi_{\mathbf{k}}^{\prime}+\xi_{\mathbf{k}%
+\mathbf{Q}}^{\prime}}{2}\nonumber\\
&  \pm\sqrt{\left(  \frac{\xi_{\mathbf{k}}^{\prime}-\xi_{\mathbf{k}%
+\mathbf{Q}}^{\prime}}{2}\right)  ^{2}+U_{ff}^{2}\left\vert \Delta
_{SDW}\right\vert ^{2}}.
\end{align}
The same set of equations become the ones for the CDW state with the
substitution of $\Delta_{SDW}$ with $\Delta_{CDW}$ and $U_{ff}$ with $-\left[
2U_{ind}\left(  \mathbf{Q}\right)  +U_{ff}\right]  $. \ \ In the weak
interacting regime where one can ignore the Fock potential and fix the
chemical potential according to the number equation (\ref{nf}) for a normal
gas at $T=0$, \ the critical temperatures for the SDW and CDW phase are
determined from the equation,%
\begin{equation}
s\left(  \bar{T}\right)  =1,
\quad c\left(  \bar{T}\right)  =1,
\end{equation}
where
\begin{align}
s  &  =\frac{\bar{U}_{ff}}{\pi^{2}}\int_{0}^{1}P\left(  x,\bar{T}\right)
K\left(  1-x^{2}\right)  ,\label{sdw}\\
c  &  =-\frac{2\bar{U}_{ind}\left(  \mathbf{Q}\right)  +\bar{U}_{ff}}{\pi^{2}%
}\int_{0}^{1}P\left(  x,\bar{T}\right)  K\left(  1-x^{2}\right)  , \label{cdw}%
\end{align}
with
\begin{equation}
P\left(  x,\bar{T}\right)  =\frac{1}{x}\left(  \tanh\frac{x-\bar{\mu}_{f}%
}{2\bar{T}}+\tanh\frac{x+\bar{\mu}_{f}}{2\bar{T}}\right)  ,
\end{equation}
and
\begin{equation}
\bar{U}_{ind}\left(  \mathbf{Q}\right)  =-\frac{\bar{U}_{0}}{4\xi^{2}b\left(
1+\chi^{-1}\right)  },
\end{equation}
where $\bar{U}_{ind}\left(  \mathbf{Q}\right)  \equiv U_{ind}\left(
\mathbf{Q}\right)  /4t_{f}$ is the scaled induced interaction at the nesting wavevector.

Before presenting numerical results in the next section, a comment is in order.  Our results for the onset of superfluid pairings and density waves was derived within the usual mean-field approach, in which superfluid instability is treated independently from density-wave instability.  A more rigorous treatment for systems operating at (or near) half filling is the parquet method \cite{dzyaloshinskii88JETP.94.344}, which couples the two instabilities.  The parquet method calculates the sum of an infinite series of diagrams using flow equations from1 renormalization group theory, a system of nonlinear differential equations.  In 1D near half filling, the parquet prediction is drastically different from the mean-field prediction, but in 2D near half filling the predictions qualitatively agree \cite{zheleznyak97PhysRevB.55.3200}.  A quantitative comparison of the two approaches requires adapting the parquet method to our 2D model, which is beyond the scope of the current work and which we leave for future study.

\section{Numerical Results and Discussions}

In this section, we apply the formulas outlined in previous sections to
numerically determine critical temperatures and use these results as a guide
to understanding the phase structures and competing orders of the system.

\subsection{Parameters}

\begin{figure}
[ptb]
\begin{center}
\includegraphics[
width=3.4in
]%
{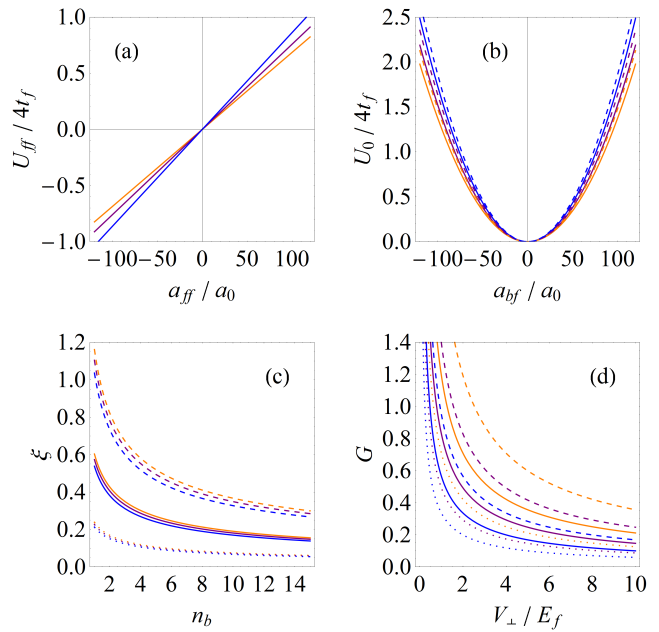}%
\caption{(Color online) (a) $U_{ff}/4t_{f}$ as a function of $a_{ff}$, (b)
$U_{0}/4t_{f}$ [defined in Eq. (\ref{Uind})] as a function of $a_{bf}$, (c)
$\xi$ as a function of $n_{b}$, and (d) $G$ [defined in Eq. (\ref{G})] as a
function of $V_{\perp}/E_{f}$. In all diagrams, $\lambda=1060$ nm, and
$V_{z}/V_{\perp}=4$ (orange), $6$ (violet)$,$ and $10$ (blue). $V_{\perp
}/E_{r}=5$ in (a), (b) and (c), and $a_{bb}=50$ $a_{0}$ in (b) and (c), where
$a_{0}$ is the Bohr radius. \ In (a), there exists only one group of lines
because $U_{ff}/4t_{f}$ is mass independent. \ In (b), the solid curves are
for $m_{f}/m_{b}=1$ while the dashed curves are for both $m_{f}/m_{b}=2$ and
$m_{f}/m_{b}=1/2$ because $U_{0}/4t_{f}$ is invariant with respect to the
substitution $m_{f}/m_{b}\rightarrow m_{b}/m_{f}$. \ In (c) and (d), the solid
ones are for $m_{f}/m_{b}=1,$ the dashed ones are for $m_{f}/m_{b}=2$, and the
dotted ones are for $m_{f}/m_{b}=1/2$. \ }%
\label{Fig:parameters}%
\end{center}
\end{figure}

The critical temperatures depend on several parameters. \ Consider first
$\bar{U}_{ff}$, $\bar{U}_{0}$, and $\xi$,\ which are functions of and thus
controlled by $a_{ff}$, $a_{bf}$, and $n_{b}$. \ In Fig. \ref{Fig:parameters},
we plot (a) $\bar{U}_{ff}$ as a function of $a_{ff}$ under different ratios of
the axial to radial trap potential, and (b) $\bar{U}_{0}$ as a function of
$a_{bf}$ and (c) $\xi$ as a function of $n_{b}$ for different ratios of the
axial to radial trap potentials as well as different ratios of the fermion to
boson mass. \ Here and throughout what follows, we fix the wavelength to
$\lambda=1060$ nm and assume that $V_{f\perp}=V_{b\perp}\equiv V_{\perp}$ and
$V_{fz}=V_{bz}\equiv V_{z}$. \ It is to be stressed that we do not intend to
limit our investigation to particular systems in view of the rich existence of
atomic elements and their isotopes in nature; the particular choices here in
Fig. \ref{Fig:parameters} are nothing more than to establish the orders of
magnitude for the ranges in which $\bar{U}_{ff}$, $\bar{U}_{0}$, and $\xi$
can vary when the corresponding control parameters are tuned within some
realistic domains. \ As can be seen, they do not change significantly with the
ratio of the axial to radial trap potential (so long as they are within the
quasi-2D regime), but the healing length $\xi$ is sensitive to the
fermion-boson mass ratio and the larger this ratio, the higher the healing
length.%

The next parameter in line is $\chi_{dd}$, defined in Eq. (\ref{chi_dd}), a
control knob inaccessible to nondipolar Bose-Fermi mixtures. \ As can be seen,
by adjusting the polar angle of the dipole $\theta_{d}$ and the dipolar
interaction $\varepsilon_{dd}$, one can change $\chi_{dd}$ continuously from
positive values where the dipole is oriented close to the $z$ axis with
$\theta_{d}<\sin^{-1}\sqrt{2/3}=54.7^{%
{{}^\circ}%
}$ to negative ones where the dipole is tilted away from the $z$ axis with
$\theta_{d}>54.7^{%
{{}^\circ}%
}$. \ Finally, we consider $G$ defined in Eq. (\ref{b}). \ It can be shown
that $G$ is determined by the trap geometry according to%
\begin{equation}
G=6\sqrt{\frac{\pi}{2}}\frac{1}{\zeta}\frac{1}{\pi^{3}}\left(  \frac{m_{f}%
}{m_{b}}\right)  ^{3/4}\left(  \frac{E_{f}}{V_{bz}}\right)  ^{1/4}\left(
\frac{E_{f}}{V_{b\perp}}\right)  ^{1/2}, \label{G}%
\end{equation}
where $\zeta$ has been defined previously in Eq. (\ref{beta}). \ As can be
seen from Eq. (\ref{G}), $G$ becomes small when $V_{bz}$ and $V_{b\perp}$
become large in comparison with the photon recoil energy $E_{f}$. \ Thus, it
is not surprising that for a quasi-2D trap where $V_{bz}\gg E_{f}$, $G$ is
found to be less than 1 for typical values of $V_{b\perp}$, which for our
interests lie within $\left(  3\sim6\right)  E_{f}$ as illustrated in Fig.
\ref{Fig:parameters}(d). \ From now on, without loss of generality, we fix $G$
to $0.3754$ for all calculations.

\begin{figure}
[ptb]
\begin{center}
\includegraphics[
width=2.5in
]%
{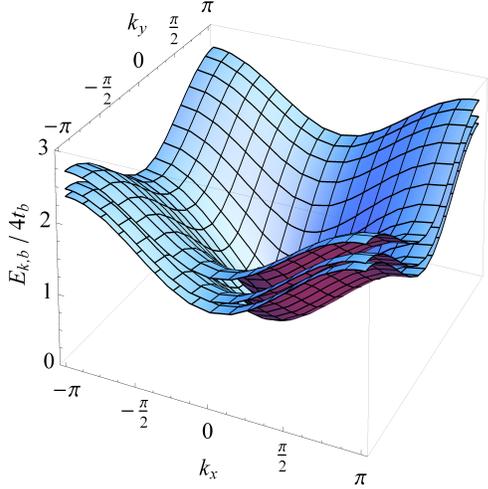}%
\caption{{}(Color online) The Bogoliubov phonon spectra, $E_{\mathbf{k}%
,b}/4t_{f}=\sqrt{\left(  4-\gamma_{\mathbf{k}}\right)  \left[  1+2\chi
_{dd}\left(  1+G\gamma_{\mathbf{k}}\right)  +\xi^{2}\left(  4-\gamma
_{\mathbf{k}}\right)  \right]  }/4\xi$ (which is Eq. (\ref{Ekb}) for an
isotropic NN interaction), for $\xi = 0.5$ and $\varepsilon_{dd}=0.5$ (top), $\varepsilon
_{dd}=0.0$ (middle) and $\varepsilon_{dd}=-0.2$ (bottom). For all diagrams that follow in this paper, $G=0.3754$.}%
\label{Fig:phononSpectrum}%
\end{center}
\end{figure}

For systems with $G<1$, a straightforward analysis indicates that the phonon
spectrum near $\mathbf{k}=0$ acquires an imaginary component when $\chi_{dd}$
$<\chi_{dd}^{th}\equiv-1/[2\left(  1+G\right)  ]$ \ and as a result, only when
$\chi_{dd}>\chi_{dd}^{th}$, does the homogeneous dipolar BEC become stable
against collapse. \ Figure \ref{Fig:phononSpectrum} displays, within the first
Brillouin zone, the phonon dispersion spectra for different dipolar
interaction $\chi_{dd}$. \ The middle surface corresponds to the spectrum
without the dipolar interaction. \ Evident in Fig. \ref{Fig:phononSpectrum} is
that increasing $\left\vert \chi_{dd}\right\vert $ on the positive side (top
surface) raises the phonon energy relative to the middle one. Contrarily
increasing $\left\vert \chi_{dd}\right\vert $ on the negative side (bottom
surface) lowers the phonon energy. \ This latter ability provides us with the
opportunity to enhance the phonon-induced Fermi-Fermi interaction by tuning
$\chi_{dd}$ on the negative side (but keeping it less than $\chi_{dd}^{th}$).%
\ \ \ 

\subsection{Pairing-Specific \textquotedblleft Effective\textquotedblright%
~Interactions\ }

Having discussed the key parameters, we now seek to gain qualitative
understanding of the critical temperatures as functions of these parameters.
\ We do so from \textquotedblleft effective\textquotedblright\ interactions
which we now define. The critical temperatures are determined by matrix
elements such as those in Eq. (\ref{s}), each of which can be decomposed into
a 1D integral, which is a function of $T$ and $\mu_{f}$ (and is always
positive), and a coefficient such as $I_{2}$ in Eq. (\ref{I2}), which is a
function of the key parameters just discussed above. \ This latter quantity is
what we regard as the \textquotedblleft effective\textquotedblright%
\ interaction strength for the corresponding phase where the symmetry of the
phase has been taken into consideration. There is a total of seven phases,
which we divide into two categories. The first one consists of $d_{x^{2}%
-y^{2}}$-, $d_{xy}$-, $g$-, and $p$-wave pairing phases, whose critical
temperatures are independent of $U_{ff}$. \ The second one consists of the
$s$-wave pairing state, SDW ordering, and CDW ordering, whose critical
temperatures can be either enhanced or suppressed by adjusting $U_{ff}$.\ \ 

\begin{figure}
[ptb]
\begin{center}
\includegraphics[
width=3.35in
]%
{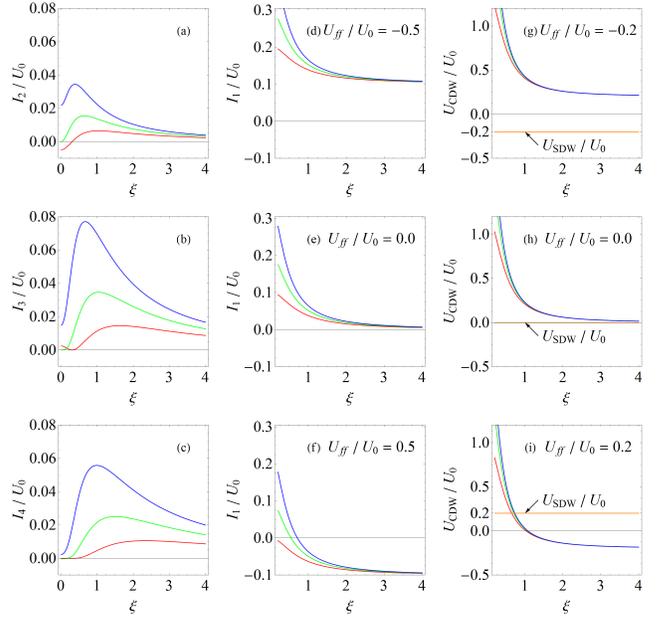}%
\caption{(Color online) Column 1: (a) $I_{2}/U_{0}$ [Eq. (\ref{I2})], (b)
$I_{3}/U_{0}$ [Eq. (\ref{I3})], and (c) $I_{4}/U_{0}$ [Eq. (\ref{I4})] as a
function of the healing length $\xi$. Column 2: $I_{1}/U_{0}$ [Eq. (\ref{I1})]
as a function of $\xi$ when $U_{ff}/U_{0}=0.5$ (e), $0.0$ (f), and $-0.5$
(g). \ Column 3: $U_{CDW}\left(  \equiv-2U_{ind}(\mathbf{Q}\right)
-U_{ff})/U_{0}$ and $U_{CDW}\left(  \equiv U_{ff}\right)  /U_{0}$ as functions
of $\xi$ when $U_{ff}/U_{0}=0.2$ (h), $0.0$ (i), and $-0.2$ (g). \ Each figure
in columns 1 and 2 consists of three curves corresponding to, from\textit{
bottom} to \textit{top}, $\varepsilon_{dd}=0.5,~0$, and $-0.2$. }%
\label{Fig:coefficients}%
\end{center}
\end{figure}

In the first column of Fig. \ref{Fig:coefficients}, we plot the ratios,
$I_{2}/U_{0}$, $I_{3}/U_{0}$, and $I_{4}/U_{0}$, which are responsible for the
phases in the first category, as functions of the healing length $\xi$ under
different dipolar interactions $\chi_{dd}$. \ Let us first turn our attention
to the middle curve in each plot in the first column, which is produced in the
absence of the dipolar interaction ($\chi_{dd}=0$). \ (a) $I_{2}$, (b) $I_{3}%
$, and (c) $I_{4}$ reach their peak when $\xi$ is set at $0.64$, $1.04,$ and
$1.51$, respectively. \ This implies that in the weak coupling limit, the
corresponding critical temperature is expected to reach its maximum at the
corresponding peak healing length, independent of $U_{0}$. \ It is interesting
to observe that the maximum healing length $\xi=$ $0.64$ for $I_{2}$, which we
obtain here in a semi-analytical fashion, almost equals to what Wang\textit{
et al.} \cite{wang05PhysRevA.72.051604} found from their numerical analysis of
the critical temperature for the $d_{x^{2}-y^{2}}$-wave pairing. \ Next, we
turn to the curves produced using non-zero $\chi_{dd}$. \ The bottom curves of
Fig. \ref{Fig:coefficients}(a), (b) and (c) in the first column indicate that
increasing the dipolar interaction on the positive side reduces the peak value
and shifts the peak towards high healing length while the top curves show
that increasing the dipolar interaction on the negative side increases the
peak value and pushes the peak position towards low healing lengths. Thus,
in our study of the critical temperature in the next section, we will always
use negative $\chi_{dd}$.%
\ \ 

Let us now turn our attention to the \textquotedblleft
effective\textquotedblright\ interactions for the phases in the second
category. \ Consider, first, the s-wave pairing, which, in general, is a
coherent superposition between the onsite and extended s-wave pairings. \ Its
critical temperature, the root to Eq. (\ref{s-wave superfluid}), depends on
the interplay between the $I_{1}$ and $I_{2}$ coefficients in a non-intuitive
fashion. \ It is thus quite difficult to single out, under general
circumstances, the \textquotedblleft effective\textquotedblright\ interaction
responsible for the s-wave pairing. However, in the case of half filling, the
off diagonal elements, $s_{0}^{1}$ and $s_{1}^{0}$, vanish so that Eq.
(\ref{s-wave superfluid}) reduces to the diagonal form
\begin{equation}
\det\left\vert
\begin{array}
[c]{cc}%
s_{0}^{0}\left(  \bar{T}\right)  -1 & 0\\
0 & s_{1}^{1}\left(  \bar{T}\right)  -1
\end{array}
\right\vert =0,
\end{equation}
which allows us to identify, straightforwardly, that the ``effective" interactions are $I_{1}$ for the onsite
s-wave pairing and $I_{2}$ for the extended s-wave pairing. \ As
illustrated in the plots in the second column of Fig. \ref{Fig:coefficients},
\ in contrast to $I_{2}$, $I_{3}$, and $I_{4}$, the coefficient $I_{1}$ is a
monotonically decreasing function of $\xi$. \ \ This is nothing more than a
simple reflection of the well established fact that fermions favor the s-wave
pairing more than unconventional pairings in the limit of a short healing
length. \ As expected, in comparison with Fig. \ref{Fig:coefficients}(e) where
$U_{ff}=0$, attractive $U_{ff}$ enhances the onsite s-wave pairing [Fig.
\ref{Fig:coefficients}(f)] while repulsive $U_{ff}$ suppresses or even
excludes (in the region where $I_{1}$ is negative) the onsite s-wave pairing
[Fig. \ref{Fig:coefficients}(d)].

In addition to the s-wave pairing, the second category also contains the SDW
and CDW ordering. \ In the SDW state, fermions with opposite spins tend to be
distributed in different sites, which is only possible for repulsive fermions.
The phonon-induced interaction is always attractive. \ This explains why the
effective interaction for the SDW phase, $U_{SDW}\equiv U_{ff}$, is completely
determined by the s-wave interaction between fermions of opposite spins
$U_{ff}$ [Eq. (\ref{sdw})] and exists only when $U_{ff}>0$. \ In the CDW
phase, fermions of opposite spins tend to aggregate on the same site, \ which
is only possible for attractive fermions. \ \ The phonon-induced interaction
is thus the main source of the interaction responsible for the CDW phase. \ In
addition, as shown in the last column of Fig. \ref{Fig:coefficients}, compared
to Fig. \ref{Fig:coefficients}(h) where $U_{ff}=0$, attractive $U_{ff}$ [Fig.
\ref{Fig:coefficients}(g)] shall help increase the effective interaction for
the CDW\ phase, $U_{CDW}$ $\equiv-2U_{ind}\left(  \mathbf{Q}\right)  -U_{ff}$,
and hence enhance CDW, while repulsive $U_{ff}$ [Fig. \ref{Fig:coefficients}%
(i)] shall help decrease $U_{CDW}$ and hence suppress CDW. \ In Fig.
\ref{Fig:coefficients}(i), CDW does not exist in the region where $U_{CDW}<0$,
and dominates SDW in the region on the left side of the intercept between
$U_{CDW}$ and $U_{SDW}$ in which $-U_{ind}\left(  \mathbf{Q}\right)  =U_{ff}$.%
\begin{figure}
[ptb]
\begin{center}
\includegraphics[
width=3.3in
]%
{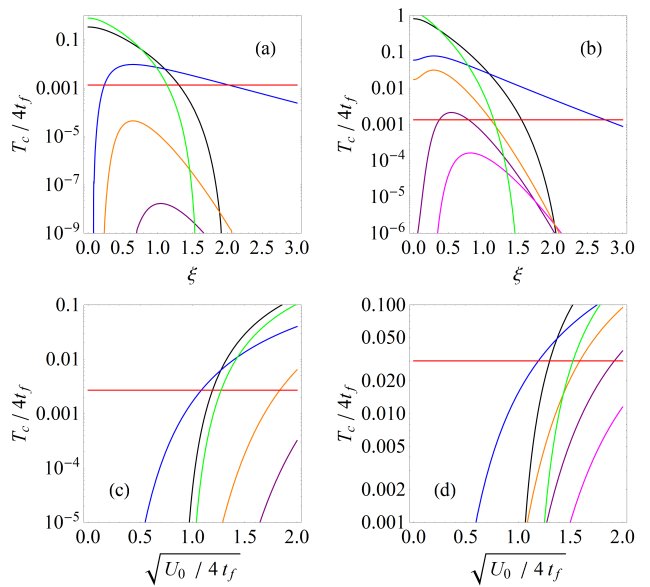}%
\caption{(Color online) $T_{c}$ as functions of the healing length $\xi$ at
half filling for (a) $\chi_{dd}=0$ and (b) $\chi_{dd}=-0.25$ when $\sqrt
{U_{0}/4t_{f}}=1.3$ and $U_{ff}/4t_{f}=0.15$. \ $T_{c}$ as functions of
$\sqrt{U_{0}/4t_{f}}$ at half filling for (c) $\chi_{dd}=0$, $\xi=1.0$, and
$U_{ff}/4t_{f}=0.18$ and (d) $\chi_{dd}=-0.25$, $\xi=0.8$, and $U_{ff}%
/4t_{f}=0.4$. Throughout this paper, we adopt the following color code scheme
for $T_{c}$: \ black for s-wave, blue for $d_{x^{2}-y^{2}}$-wave, purple for
$d_{xy}$-wave, magenta for $g$-wave, and orange for $p$-wave superfluid
phases, and red for SDW and green for CDW orderings. Furthermore, for the
phases of missing colors, it is implied that they either do not exist or have a
$T_{c}$ below the temperature scale of the relevant diagram.}%
\label{Fig:Tc1}%
\end{center}
\end{figure}

\subsection{Critical Temperatures}

Figures \ref{Fig:Tc1}(a) and (b) show $T_{c}$ as a function of the healing
length $\xi$ at half filling. \ The most prominent feature is that when
$\chi_{dd}$ is tuned negative in Fig. \ref{Fig:Tc1}(b), the critical
temperatures for unconventional pairings become significantly higher than
their counterparts in Fig. \ref{Fig:Tc1}(a) where $\chi_{dd}=0$; even the
$g$-wave pairing, which is absent in Fig. \ref{Fig:Tc1}(a) for the
temperatures plotted, enters Fig. \ref{Fig:Tc1}(b). The features exhibited in
Fig. \ref{Fig:Tc1} are precisely what we expect based on the\textquotedblleft
effective\textquotedblright\ interaction analysis summarized in Fig.
\ref{Fig:coefficients}, examples of which include that both $d_{x^{2}-y^{2}}$-
and $p$-wave pairings peak at the same healing length $\sim0.6$ since both are
determined by $I_{2}$, that there is an intercept between the SDW and CDW
phase (which takes place at $-U_{ind}\left(  \mathbf{Q}\right)  =U_{ff}$), and
that the peaks shift toward the lower healing length when a negative
$\chi_{dd}$ is turned on.

Figures \ref{Fig:Tc1}(c) and (d) display $T_{c}$, again at half filling, as a
function of $\sqrt{U_{0}}$. \ As $U_{0}$ increases, the system, shown in Fig.
7(c) where $\chi_{dd}=0$, changes its phase from SDW to the $d_{x^{2}-y^{2}}%
$-wave pairing, and then to the s-wave pairing. \ There is a substantial
increase in $T_{c}$ for pairings different from $d_{x^{2}-y^{2}}$- and
$s$-wave pairings when $\chi_{dd}$ is tuned negative in Fig. \ref{Fig:Tc1}(d),
but this increase in $T_{c}$ will not alter the sequence of phases exhibited
in Fig. \ref{Fig:Tc1}(c). \ The conclusion is that at half filling, the
$d_{x^{2}-y^{2}}$-wave pairing always dominates all other unconventional
pairings in spite of the fact that the critical temperatures for the latter
may increase more dramatically than the $d_{x^{2}-y^{2}}$-wave pairing in
response to an increase in $\left\vert \chi_{dd}\right\vert $ and $U_{0}$.%
\begin{figure}
[ptb]
\begin{center}
\includegraphics[
width=3.3in
]%
{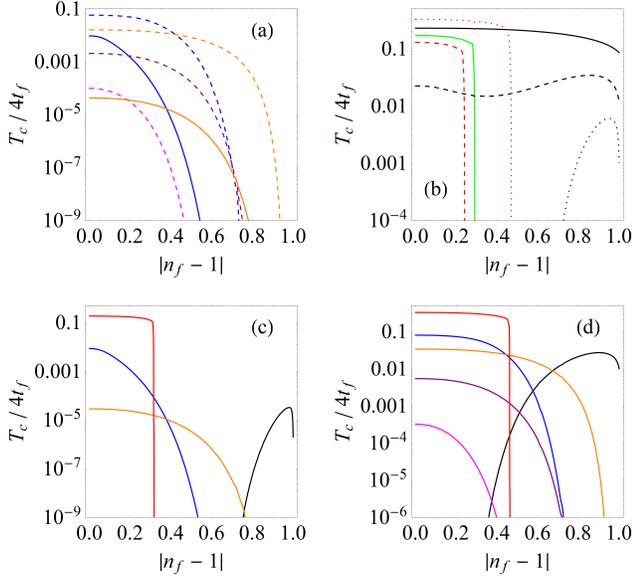}%
\caption{(Color online) $T_{c}$ as functions of $\left\vert n_{f}-1\right\vert
$. For (a) and (b), $\xi=$ $0.6$ and $\sqrt{U_{0}/4t_{f}}=1.3$. \ (a) is for
the phases in the first group (with $d_{x^{2}-y^{2}}$-$,$ $d_{xy}$-$,$ $p$-$,$
and $g$-wave pairings) with the solid and dahsed lines representing those when
$\chi_{dd}=0$ and $\chi_{dd}=-0.25$, respectively. \ (b) is for the phases in
the second group (with $s$-wave symmetry and SDW and CDW ordering) when
$\chi_{dd}=-0.25$ and $U_{ff}=0$ (solid lines), 0.8 (dashed lines), and 1.5
(dotted lines). \ For (c) and (d), $\xi=$ $0.5$ and $\sqrt{U_{0}/4t_{f}}=1.3$.
\ (c) is for the phases when $\chi_{dd}=0$ and $U_{ff}/4t_{f}=1.0$ and (d) is
for the phases when $\chi_{dd}=-0.27$ and $U_{ff}/4t_{f}=1.5$. \ These plots
are symmetric with respect to $n_{f}-1\rightarrow1-n_{f}$ due to the
particle-hole symmetry.}%
\label{Fig:Tc-nf}%
\end{center}
\end{figure}

Finally, we consider the effect of the filling factor $n_{f}$ on $T_{c}$ for
various competing orders. \ In Fig. \ref{Fig:Tc-nf}(a), we display $T_{c}$ as
a function of $\left\vert n_{f}-1\right\vert $ for the phases in the first
category when $\chi_{dd}=0$ (solid lines) and $\chi_{dd}=-0.25$ (dashed
lines). It can be seen from solid lines$~$($\chi_{dd}=0$), the $p$-wave
pairing dominates the $d_{x^{2}-y^{2}}$-wave pairing away from half filling in
the region where $\left\vert n_{f}-1\right\vert $ exceeds around 0.4. \ In the
presence of $\chi_{dd}$ (= $-0.25$) (dashed lines), the crossing from the $p$-
to $d_{x^{2}-y^{2}}$-wave pairing shifts (slightly) towards a higher
$\left\vert n_{f}-1\right\vert $ while the critical temperature for the
$p$-wave pairing experiences a significant increase. \ Note that since both
$p$-wave [Eq. (\ref{p integrals})] and $d_{x^{2}-y^{2}}$-wave [Eq. (\ref{dd})]
pairings share the same effective interaction strength $I_{2}$, in the limit
of weak interaction, the $p$- to $d_{x^{2}-y^{2}}$-wave crossing becomes
independent of $I_{2}$ and hence is fixed essentially by the pairing
symmetries. This is to be contrasted to other crossings such as from
$d_{x^{2}-y^{2}}$- to $d_{xy}$-wave as shown by the dashed lines, which
depends on the ratio of $I_{2}$ to $I_{3}$ and hence is highly susceptible to
changes in system parameters.

The system is not in the $p$-wave state unless the $p$-wave pairing also has a
higher $T_{c}$ than the phases in the second category. \ In Fig.
\ref{Fig:Tc-nf}(b), we show $T_{c}$ as a function of $\left\vert
n_{f}-1\right\vert $ for the phases in the second category under different
$U_{ff}$ $\left(  \geq0\right)  $ with the solid lines for $U_{ff}=0$, the
dashed lines for $U_{ff}=$ $0.8$, and the dotted lines for $U_{ff}=1.5$. \ In
the absence of $U_{ff}$ (solid lines), SDW does not exist due to the lack of
repulsive interaction, CDW (solid green line) exists close to half filling due
to the phonon-induced attraction, and the s-wave pairing (solid black line)
has a high $T_{c}$ throughout. \ Increasing $U_{ff}$ suppresses CDW (not shown
due to a low temperature) while enhancing and expanding SDW (dashed and dotted
red lines), but most importantly it suppresses the s-wave pairing in the
region of low $\left\vert n_{f}-1\right\vert $ (dashed and dotted black
lines). \ This suppression is strong enough that, as shown by the dotted lines
in \ref{Fig:Tc-nf}(b), all phases in the second category are strongly
suppressed in the same region where the $p$-wave pairing exceeds the
$d_{x^{2}-y^{2}}$-wave pairing displayed in Fig. \ref{Fig:Tc-nf}(a).

This, in turn, creates a window of opportunity for the $p$-wave superfluid to
come out a winner. \ Two such examples are shown, respectively, in Fig
\ref{Fig:Tc-nf}(c) and (d) with and without dipolar interaction. \ In
particular, Fig \ref{Fig:Tc-nf}(d) demonstrates that the $p$-wave pairing with
a critical temperature in the order of a hundredth of $4t_{f}$ is possible.
\ The conclusion is that in contrast to the $d_{x^{2}-y^{2}}$-wave pairing,
which may dominate close to half filling, the $p$-wave may dominate away from
half filling in a window centered roughly around $\left\vert n_{f}%
-1\right\vert =0.6$, \ while all the other unconventional pairings, in spite
of the possibility of being substantially enhanced by the use of the dipolar
interaction, are unable to emerge as dominant phases. \ 

\section{Conclusion}

We have considered a cold atom mixture in a square optical lattice where the
mixture is made up of a two-component Fermi gas and a single-component dipolar
Bose gas in the state of a homogeneous BEC. \ By eliminating the phonons of
the dipolar BEC, we have constructed an effective lattice Fermi system which
preserves the symmetry of point group $D_{4}$. \ Focusing on this effective
model, we have explored, within the Hartree-Fock-Bogoliubov mean-field theory,
the competition among density waves and superfluids with both conventional and
unconventional parings. \ We have constructed, in the
weak coupling regime, the matrix representation of the linearized gap equation
in the irreducible representations of $D_{4}$. \ We have simplified each
matrix element from a 4D integral into a separable form involving a 1D
integral, which is only a function of $T$ and $\mu_{f}$, and a
pairing-specific \textquotedblleft effective\textquotedblright\ interaction,
which is an analytical function of the parameters that characterize the
Fermi-Fermi interactions in our system. 

To help appreciate the significance of these 1D integral equations, we
remind that for a continuous 3D two-component Fermi gas, there exists a
similar 1D integral equation
\cite{stoof09UltracoldQuantumFieldsBook} 
\begin{equation}
-\frac{2k_{F}a_{s}}{\pi}\int_{0}^{\infty}\sqrt{x}\left[  \frac{\tanh
\frac{x-\bar{\mu}_{f}}{2\bar{T}}}{2\left(  x-\bar{\mu}_{f}\right)  }-\frac
{1}{2x}\right]  dx=1,\label{3D S}%
\end{equation}
which is the origin of the well-known formula, 
\begin{equation}
\bar{T}\approx8e^{\gamma-2}e^{-\pi/2k_{F}a_{s}}/\pi
,\label{3D critical temperature}%
\end{equation}
that one uses to estimate the critical temperature $\bar{T}$
of s-wave superfluid pairing, where $k_{F}$ is the Fermi wave number,
$a_{s}$ the s-wave scattering length, and $\gamma$
Euler-Mascheroni constant.

The 1D integral equations we highlighted in Sec. IV are analogs of Eq.
(\ref{3D S}) for\ fermions in a square lattice that interact via both a
contact and a long-range interaction. \ The long-range interaction here is in
the form of an attractive (lattice-version) Yukawa potential, which occurs
ubiquitously across a broad spectrum of physics, e.g., nuclear physics and
condensed matter physics. Thus, in analogy to Eq. (\ref{3D S}), we expect that
our 1D equations will find applications beyond the current model of
interest.\ Further, the ability to reduce our gap equations into
ones\ analogous to Eq. (\ref{3D S}) may hint that similar reductions may exist
for interaction potentials with mathematical forms different from the Yakawa type.  The various techniques we developed in this work may then offer a good starting point for future studies to build toolboxes for simplifying gap equations associated with such interactions.

As Eq. (\ref{3D S}), our 1D integral equations enjoy the advantage of
allowing many features of the critical temperatures to be determined in a
semi-analytical manner. In particular, we have applied them to
analyze the critical temperatures of various competing orders as functions of
the healing length and filling factor in both the absence and presence of the
dipolar interaction. \ We have found that tuning dipolar interaction
$\chi_{dd}$ on the negative side lowers the phonon energy and can
significantly enhance the unconventional pairings. We have found that close to
half filling, the $d_{x^{2}-y^{2}}$-wave pairing with a critical temperature
in the order of a fraction of $4t_{f}$ may dominate all the other phases, and
at a higher filling factor, the $p$-wave pairing with a critical temperature
in the order of a hundredth of $4t_{f}$ may emerge as a winner. Our theory
also includes the pairings with $d_{xy}$- and $g$- wave symmetries. In spite
of dramatic enhancements of their critical temperatures, we have found that
tuning a dipolar interaction will not be able to make the pairings with
$d_{xy}$- and $g$- wave symmetries to dominate the other phases. 

It is to be stressed that the phase diagram based on the critical
temperature analysis does not correspond necessarily to what is realizable at
zero temperature. \ The actual ground state may exist as a superfluid with a
pure symmetry or a coherent superposition of those with pure symmetries.
\ [The latter dominates the phase diagram when the dipolar interaction is
tuned away from the diagonal line (but stays within the square) in Fig.
\ref{Fig:regionOfDipolarBEC} where the system breaks D4 symmetry.] \ \ To
construct the phase diagram at zero temperature requires the computational
machinery of energy minimization, which can be a time-consuming task for a
dipolar Bose-Fermi mixture where the parameter space is unusually large. \ The
1D integrals developed in this paper allow us to gain not only quick insights
into various critical temperatures but also, very importantly, a sense for the
role each parameter plays in this large parameter space. Thus, these
integrals, just as with their 3D continuous analog in Eq. (\ref{3D S}), serve as a
valuable theoretical tool that one can use to perform a preliminary analysis
before embarking on more complex tasks such as energy minimization.

\section*{Acknowledgments}

B.\ K.\ is grateful to ITAMP and the Harvard-Smithsonian Center for Astrophysics for their hospitality while completing this work.  H.\ Y.\ L.\ was supported in part by the US Army Research Office under Grant No.\
W911NF-10-1-0096 and in part by the US National Science Foundation under Grant
No.\ PHY11-25915.

\appendix
\section{Appendix}

In this appendix, we outline the main steps that we use to reduce $d_{0}$, the
matrix element for the d-wave pairing, from a 4D integral in Eq. (\ref{d0}) to
a 1D integral in Eq. (\ref{dd}), and use it as an example to showcase the
techniques that we employed to simplify the matrix elements for all other
pairings from 4D integrals in Eqs. (\ref{s}), (\ref{d}) and (\ref{triplet}) to
1D integrals in Eqs. (\ref{s integtrals}), (\ref{p integrals}), (\ref{ee}),
and (\ref{gg}). To begin with, we organize Eq. (\ref{d0}), a 4D integral over
the first Brilliouin zone in momentum space, into a 2D integral involving
$S_{\mathbf{q}}\left(  T\right)  $
\begin{equation}
d_{0}=\frac{1}{4\left(  2\pi\right)  ^{4}}\int\int L_{\mathbf{q}}%
S_{\mathbf{q}}\left(  T\right)  \eta_{\mathbf{q}}d^{2}\mathbf{q,} \label{d0 a}%
\end{equation}
in terms of another 2D integral involving the Fermi-Fermi interaction
$U_{s}\left(  \mathbf{k,q}\right)  $%
\begin{equation}
L_{\mathbf{q}}=-\int\int U_{s}\left(  \mathbf{k,q}\right)  \eta_{\mathbf{k}%
}d^{2}\mathbf{k.} \label{Kd}%
\end{equation}
As a first step to solving Eq. (\ref{Kd}), we\ substitute $U_{s}\left(
\mathbf{k,q}\right)  $ in Eq. (\ref{Us}), where $U_{ind}\left(  \mathbf{k}%
\right)  $ is given by Eq. (\ref{Uind 1}), into Eq. (\ref{Kd}), and change Eq.
(\ref{Kd}) into
\begin{equation}
L_{\mathbf{q}}=U_{0}\int\int\frac{\eta_{\mathbf{k}}d^{2}\mathbf{k}}%
{1+2\chi_{dd}+\xi^{2}\left(  4-b\gamma_{\mathbf{k}-\mathbf{q}}\right)  }
\label{Kai d}%
\end{equation}
where we have used the fact that the integral involving $U_{ff}$ is zero and
the one involving $U_{ind}\left(  \mathbf{k}+\mathbf{q}\right)  $ is the same
as the one involving $U_{ind}\left(  \mathbf{k}-\mathbf{q}\right)  $. \ For
notational simplicity, we put Eq. (\ref{Kai d}) in a more compact form
\[
L_{\mathbf{q}}=\tilde{U}_{0}\int\int\frac{\eta_{\mathbf{k}}d^{2}\mathbf{k}%
}{1+\tilde{\xi}^{2}\left(  4-b\gamma_{\mathbf{k}-\mathbf{q}}\right)  }%
\]
in terms of $\tilde{U}_{0}$ and $\tilde{\xi}$ defined as
\begin{equation}
\tilde{U}_{0}=\frac{U_{0}}{1+2\chi_{dd}},
\quad
\tilde{\xi}=\frac{\xi}{\sqrt
{1+2\chi_{dd}}}. \label{new U_0 and ksi}%
\end{equation}
Next, we make the change of variables $\mathbf{k}\rightarrow\mathbf{k}%
-\mathbf{q}$, and by taking advantage of the integrand being a periodic
function of momentum, we transform Eq. (\ref{Kai d}) into%
\begin{equation}
L_{\mathbf{q}}=\tilde{U}_{0}\int\int\frac{\eta_{\mathbf{k}+\mathbf{q}}%
d^{2}\mathbf{k}}{1+\tilde{\xi}^{2}\left(  4-b\gamma_{\mathbf{k}}\right)  },
\label{K q}%
\end{equation}
where $\eta_{\mathbf{k}+\mathbf{q}}=2\left[  \cos\left(  k_{x}+q_{x}\right)
-\cos\left(  k_{y}+q_{y}\right)  \right]  $ or%
\begin{align}
\eta_{\mathbf{k}+\mathbf{q}}  &  =2\left(  \cos k_{x}\cos q_{x}-\cos k_{y}\cos
q_{y}\right) \nonumber\\
&  -2\left(  \sin k_{x}\sin q_{x}-\sin k_{y}\sin q_{y}\right)  .
\label{eta k+q}%
\end{align}
By virtue of symmetry considerations, the integrals involving the sine
functions in the second line of Eq. (\ref{eta k+q}) vanish while those
involving the cosine functions in the first line of Eq. (\ref{eta k+q}) can be
put into a separable form
\begin{equation}
L_{\mathbf{q}}=\tilde{U}_{0}\frac{\eta_{\mathbf{q}}}{4}\int\int\frac
{\gamma_{\mathbf{k}}d^{2}\mathbf{k}}{1+\tilde{\xi}^{2}\left(  4-b\gamma
_{\mathbf{k}}\right)  }\text{.} \label{K q 1}%
\end{equation}
To proceed, we rewrite Eq. (\ref{K q 1}) as
\begin{equation}
L_{\mathbf{q}}=\tilde{U}_{0}\frac{\eta_{\mathbf{q}}}{4b\tilde{\xi}^{2}}\left[
\left(  1+4\tilde{\xi}^{2}\right)  L-4\pi^{2}\right]  , \label{Kq}%
\end{equation}
where $L$ is the integral
\begin{equation}
L=\int\int\frac{d^{2}\mathbf{k}}{1+\tilde{\xi}^{2}\left(  4-b\gamma
_{\mathbf{k}}\right)  }, \label{L}%
\end{equation}
which we now focus on. \ Integrating first with respect to $k_{x}$ from $-\pi$
to $+\pi$, we change Eq. (\ref{L}) into
\begin{equation}
L=\frac{\pi}{\left\vert b\right\vert \tilde{\xi}^{2}}\int_{-\pi}^{+\pi}%
\frac{dk_{y}}{\sqrt{\left(  \cos k_{y}-r_{+}\right)  \left(  \cos k_{y}%
-r_{-}\right)  }}, \label{L 1}%
\end{equation}
where%
\begin{equation}
r_{+}=\left\{
\begin{array}
[c]{c}%
\frac{1+4\tilde{\xi}^{2}+2\left\vert b\right\vert \tilde{\xi}^{2}}%
{2b\tilde{\xi}^{2}}\text{, if }b>0\\
\frac{1+4\tilde{\xi}^{2}-2\left\vert b\right\vert \tilde{\xi}^{2}}%
{2b\tilde{\xi}^{2}},\text{ if }b<0
\end{array}
\right.  , \label{r+}%
\end{equation}
and%
\begin{equation}
r_{-}=\left\{
\begin{array}
[c]{c}%
\frac{1+4\tilde{\xi}^{2}-2\left\vert b\right\vert \tilde{\xi}^{2}}%
{2b\tilde{\xi}^{2}}\text{, if }b>0\\
\frac{1+4\tilde{\xi}^{2}+2\left\vert b\right\vert \tilde{\xi}^{2}}%
{2b\tilde{\xi}^{2}},\text{ if }b<0
\end{array}
\right.  . \label{r-}%
\end{equation}
We next integrate Eq. (\ref{L 1}) and express the result in terms of the
elliptical integral as
\begin{equation}
L=\frac{\pi}{\left\vert b\right\vert \tilde{\xi}^{2}}\frac{4K\left[
\frac{2\left(  r_{+}-r_{-}\right)  }{\left(  r_{+}-1\right)  \left(
r_{-}+1\right)  }\right]  }{\sqrt{\left(  r_{+}-1\right)  \left(
r_{-}+1\right)  }}, \label{M K}%
\end{equation}
which, when $r_{+}$ and $r_{-}$ are replaced with Eq. (\ref{r+}) and Eq.
(\ref{r-}) becomes
\begin{equation}
L=2\pi\frac{4}{1+4\tilde{\xi}^{2}}K\left[  \frac{16b^{2}\tilde{\xi}^{4}%
}{\left(  1+4\tilde{\xi}^{2}\right)  ^{2}}\right]  .
\end{equation}
Finally, inserting this result into Eq. (\ref{Kq}), we find%
\begin{equation}
L_{\mathbf{q}}=2\pi^{4}4t_{f}I_{2}\eta_{\mathbf{q}},\nonumber
\end{equation}
expressed in terms of $I_{2}$ defined in Eq. (\ref{I2}), where the use of Eq.
(\ref{new U_0 and ksi}) has been made to revert back to the original
notations.%
\begin{figure}
[ptb]
\begin{center}
\includegraphics[
width=3.4in
]%
{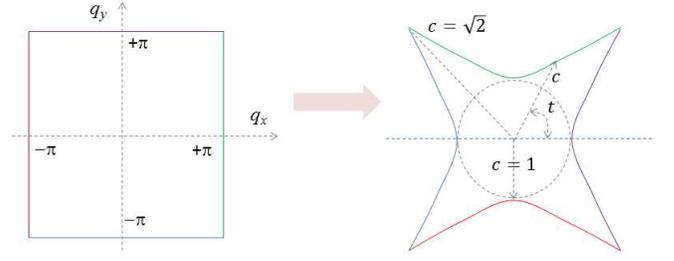}%
\caption{(Color online) A map from the first Brillouin zone in the $\left(
q_{x},q_{y}\right)  $ plane to the polar plane where $c$ is the radial
coordinate and $t$ is the angular coordinate. \ $q_{x}=-\pi$ is mapped to
$c\sin t=-1$ (red), $q_{y}=\pi$ to $c\cos t=1$ (violet), $q_{x}=+\pi$ to
$c\sin t=1$ (green), and $q_{y}=-\pi$ to $c\cos t=-1$ (blue).}%
\label{Fig:map}%
\end{center}
\end{figure}

Having solved $L_{\mathbf{q}}$, we now turn to $d_{0}$ in Eq. (\ref{d0 a}),
which, when $S_{\mathbf{q}}\left(  T\right)  $ is replaced with its scaled
form in Eq. (\ref{S(x,T)}) becomes
\begin{equation}
d_{0}=\frac{I_{2}}{2^{5}}\int\int S\left(  \bar{\epsilon}_{\mathbf{q}},\bar
{T}\right)  \eta_{\mathbf{q}}^{2}d^{2}\mathbf{q},
\end{equation}
where $\bar{\epsilon}_{\mathbf{q}}=\epsilon_{\mathbf{q},f}/4t_{f}=-\left(
\cos q_{x}+\cos q_{y}\right)  /2$. \ To solve this integral, we change
variables from $\left(  q_{x},q_{y}\right)  $ to $\left(  c,t\right)  $ via
the transformation%
\begin{equation}
c\sin t=\sin\frac{q_{x}}{2},
\quad
c\cos t=\sin\frac{q_{y}}{2},
\end{equation}
which maps the first Brillouin zone, a square in momentum space, into a
non-square region in the polar $\left(  c,t\right)  $ plane, as shown in Fig.
\ref{Fig:map} where $c$ is the radial coordinate and $t$ is the angular
coordinate. \ In terms of $c$ and $t$, we have \
\begin{equation}
d_{0}=2I_{2}\int S\left(  c^{2}-1,\bar{T}\right)  I_{t}\left(  c\right)  cdc,
\label{d0 1}%
\end{equation}
where
\begin{equation}
I_{t}\left(  c\right)  =\int\frac{\left(  c^{2}\cos^{2}t-c^{2}\sin
^{2}t\right)  ^{2}}{\sqrt{\left(  1-c^{2}\sin^{2}t\right)  \left(  1-c^{2}%
\cos^{2}t\right)  }}dt. \label{Ic(t)}%
\end{equation}
At this point, we turn to the Brillouin zone in the ($c,t$) plane in Fig.
\ref{Fig:map}. \ Due to the symmetry inherent to $I_{t}\left(  c\right)  $,
$\ I_{t}\left(  c\right)  $ is eight times the integral over the region from
$t=\pi/4$ to $\pi/2$ in the first quadrant of the polar plane, which is
divided into two areas, $\ \pi/4<t<\pi/2$ if $0<c<1$ and $\pi/4<t<\sin
^{-1}\frac{1}{c}$ if $1<c<\sqrt{2}$. Thus,
\begin{equation}
I_{t}\left(  c\right)  =\left\{
\begin{array}
[c]{c}%
\text{ }8\int_{\pi/4}^{\pi/2}\left(  \cdots\right)  dt\text{, }0<c<1,\\
8\int_{\pi/4}^{\sin^{-1}\frac{1}{c}}\left(  \cdots\right)  dt\text{,
}1<c<\sqrt{2},
\end{array}
\right.  \label{It(c) 0}%
\end{equation}
where $\left(  \cdots\right)  $ is the integrand in Eq. (\ref{Ic(t)}).
\ \ Making a change of variables from $t$ to $t^{\prime}/2+\pi/4$, we find,
with some algebraic manipulations, that
\begin{align}
I_{t}\left(  c\right)   &  =8\left(  2-c^{2}\right)  \times\nonumber\\
&  \left\{
\begin{array}
[c]{c}%
\text{ \ \ \ \ \ \ \ \ \ \ \ \ \ \ \ \ \ \ \ }K\left(  r^{2}\right)  -E\left(
r^{2}\right)  ,\text{ }0<c<1,\\
F\left(  \sin^{-1}\frac{1}{r},r^{2}\right)  -E\left(  \sin^{-1}\frac{1}%
{r},r^{2}\right)  ,\text{ }1<c<\sqrt{2},
\end{array}
\right.  \label{It(c)}%
\end{align}
where
\begin{equation}
r=c^{2}/\left(  2-c^{2}\right)  \label{r}%
\end{equation}
and $F\left(  \phi,k\right)  $ and $E\left(  \phi,k\right)  $ are the
incomplete elliptic integrals of the first and second kind, respectively.
\ Note that when $1<c<\sqrt{2}$, $r^{2}$ according to Eq. (\ref{r}) is greater
than 1 so that $F\left(  \sin^{-1}\frac{1}{r},r^{2}\right)  $ and $E\left(
\sin^{-1}\frac{1}{r},r^{2}\right)  $ in Eq. (\ref{It(c)}) are ill defined.
\ This can be circumvented by making use of the identities%
\begin{subequations}
\begin{align}
F\left[  \sin^{-1}\frac{1}{r},r^{2}\right]   &  =\frac{K\left(  r^{-2}\right)
}{r},\\
E\left(  \sin^{-1}\frac{1}{r},r^{2}\right)   &  =rE\left(  r^{-2}\right)
+\left(  \frac{1}{r}-r\right)  K\left(  r^{-2}\right)  ,
\end{align}
which can be shown to hold when $r>1$. \ With this, we simplify Eq.
(\ref{It(c)}) into
\end{subequations}
\begin{align}
I_{t}\left(  c\right)   &  =8\left(  2-c^{2}\right)  \times\nonumber\\
&  \left\{
\begin{array}
[c]{c}%
\text{ \ \ \ }K\left(  r^{2}\right)  -E\left(  r^{2}\right)  ,\text{ }0<c<1,\\
r\left[  K\left(  r^{-2}\right)  -E\left(  r^{-2}\right)  \right]  ,\text{
}1<c<\sqrt{2}.
\end{array}
\right.
\end{align}
Finally, we integrate Eq. (\ref{d0 1}) involving the piecewise function
$I_{t}\left(  c\right)  $ with respect to the radial coordinate $c,$
\begin{align}
d_{0}  &  =2I_{2}\int_{0}^{1}S\left(  c^{2}-1,\bar{T}\right)  I_{t}\left(
c\right)  cdc+\nonumber\\
&  2I_{2}\int_{1}^{\sqrt{2}}S\left(  c^{2}-1,\bar{T}\right)  I_{t}\left(
c\right)  cdc,
\end{align}
and change it into Eq. (\ref{dd}) in the main text after substituting
$c=\sqrt{1+x}$ and further algebraic manipulations.




\begin{thebibliography}{69}
\expandafter\ifx\csname natexlab\endcsname\relax\def\natexlab#1{#1}\fi
\expandafter\ifx\csname bibnamefont\endcsname\relax
  \def\bibnamefont#1{#1}\fi
\expandafter\ifx\csname bibfnamefont\endcsname\relax
  \def\bibfnamefont#1{#1}\fi
\expandafter\ifx\csname citenamefont\endcsname\relax
  \def\citenamefont#1{#1}\fi
\expandafter\ifx\csname url\endcsname\relax
  \def\url#1{\texttt{#1}}\fi
\expandafter\ifx\csname urlprefix\endcsname\relax\def\urlprefix{URL }\fi
\providecommand{\bibinfo}[2]{#2}
\providecommand{\eprint}[2][]{\url{#2}}

\bibitem[{\citenamefont{Greiner et~al.}(2002)\citenamefont{Greiner, Mandel,
  Esslinger, Hansch, and Bloch}}]{greiner02Nature.415.39}
\bibinfo{author}{\bibfnamefont{M.}~\bibnamefont{Greiner}},
  \bibinfo{author}{\bibfnamefont{O.}~\bibnamefont{Mandel}},
  \bibinfo{author}{\bibfnamefont{T.}~\bibnamefont{Esslinger}},
  \bibinfo{author}{\bibfnamefont{T.~W.} \bibnamefont{Hansch}},
  \bibnamefont{and} \bibinfo{author}{\bibfnamefont{I.}~\bibnamefont{Bloch}},
  \bibinfo{journal}{Nature} \textbf{\bibinfo{volume}{415}}, \bibinfo{pages}{39}
  (\bibinfo{year}{2002}).

\bibitem[{\citenamefont{Fisher et~al.}(1989)\citenamefont{Fisher, Weichman,
  Grinstein, and Fisher}}]{fisher89PhysRevB.40.546}
\bibinfo{author}{\bibfnamefont{M.~P.~A.} \bibnamefont{Fisher}},
  \bibinfo{author}{\bibfnamefont{P.~B.} \bibnamefont{Weichman}},
  \bibinfo{author}{\bibfnamefont{G.}~\bibnamefont{Grinstein}},
  \bibnamefont{and} \bibinfo{author}{\bibfnamefont{D.~S.}
  \bibnamefont{Fisher}}, \bibinfo{journal}{Phys. Rev. B}
  \textbf{\bibinfo{volume}{40}}, \bibinfo{pages}{546} (\bibinfo{year}{1989}).

\bibitem[{\citenamefont{Jaksch et~al.}(1998)\citenamefont{Jaksch, Bruder,
  Cirac, Gardiner, and Zoller}}]{jaksch98PhysRevLett.81.3108}
\bibinfo{author}{\bibfnamefont{D.}~\bibnamefont{Jaksch}},
  \bibinfo{author}{\bibfnamefont{C.}~\bibnamefont{Bruder}},
  \bibinfo{author}{\bibfnamefont{J.~I.} \bibnamefont{Cirac}},
  \bibinfo{author}{\bibfnamefont{C.~W.} \bibnamefont{Gardiner}},
  \bibnamefont{and} \bibinfo{author}{\bibfnamefont{P.}~\bibnamefont{Zoller}},
  \bibinfo{journal}{Phys. Rev. Lett.} \textbf{\bibinfo{volume}{81}},
  \bibinfo{pages}{3108} (\bibinfo{year}{1998}).

\bibitem[{\citenamefont{Bloch et~al.}(2008)\citenamefont{Bloch, Dalibard, and
  Zwerger}}]{bloch08RevModPhys.80.885}
\bibinfo{author}{\bibfnamefont{I.}~\bibnamefont{Bloch}},
  \bibinfo{author}{\bibfnamefont{J.}~\bibnamefont{Dalibard}}, \bibnamefont{and}
  \bibinfo{author}{\bibfnamefont{W.}~\bibnamefont{Zwerger}},
  \bibinfo{journal}{Rev. Mod. Phys.} \textbf{\bibinfo{volume}{80}},
  \bibinfo{pages}{885} (\bibinfo{year}{2008}).

\bibitem{schrieffer64Book}
J.\ R.\ Schrieffer, ``Theory of Superconductivity," New York: Benjamin (1964).

\bibitem[{\citenamefont{Bijlsma et~al.}(2000)\citenamefont{Bijlsma, Heringa,
  and Stoof}}]{bijlsma00PhysRevA.61.053601}
\bibinfo{author}{\bibfnamefont{M.~J.} \bibnamefont{Bijlsma}},
  \bibinfo{author}{\bibfnamefont{B.~A.} \bibnamefont{Heringa}},
  \bibnamefont{and} \bibinfo{author}{\bibfnamefont{H.~T.~C.}
  \bibnamefont{Stoof}}, \bibinfo{journal}{Phys. Rev. A}
  \textbf{\bibinfo{volume}{61}}, \bibinfo{pages}{053601}
  (\bibinfo{year}{2000}).

\bibitem[{\citenamefont{Viverit}(2002)}]{viverit02PhysRevA.66.023605}
\bibinfo{author}{\bibfnamefont{L.}~\bibnamefont{Viverit}},
  \bibinfo{journal}{Phys. Rev. A} \textbf{\bibinfo{volume}{66}},
  \bibinfo{pages}{023605} (\bibinfo{year}{2002}).

\bibitem[{\citenamefont{Wang et~al.}(2005)\citenamefont{Wang, Lukin, and
  Demler}}]{wang05PhysRevA.72.051604}
\bibinfo{author}{\bibfnamefont{D.-W.} \bibnamefont{Wang}},
  \bibinfo{author}{\bibfnamefont{M.~D.} \bibnamefont{Lukin}}, \bibnamefont{and}
  \bibinfo{author}{\bibfnamefont{E.}~\bibnamefont{Demler}},
  \bibinfo{journal}{Phys. Rev. A} \textbf{\bibinfo{volume}{72}},
  \bibinfo{pages}{051604} (\bibinfo{year}{2005}).

\bibitem[{\citenamefont{Truscott et~al.}(2001)\citenamefont{Truscott, Strecker,
  McAlexander, Partridge, and Hulet}}]{truscott01Science.291.2570}
\bibinfo{author}{\bibfnamefont{A.~G.} \bibnamefont{Truscott}},
  \bibinfo{author}{\bibfnamefont{K.~E.} \bibnamefont{Strecker}},
  \bibinfo{author}{\bibfnamefont{W.~I.} \bibnamefont{McAlexander}},
  \bibinfo{author}{\bibfnamefont{G.~B.} \bibnamefont{Partridge}},
  \bibnamefont{and} \bibinfo{author}{\bibfnamefont{R.~G.} \bibnamefont{Hulet}},
  \bibinfo{journal}{Science} \textbf{\bibinfo{volume}{291}},
  \bibinfo{pages}{2570} (\bibinfo{year}{2001}).

\bibitem[{\citenamefont{Schreck et~al.}(2001)\citenamefont{Schreck, Khaykovich,
  Corwin, Ferrari, Bourdel, Cubizolles, and
  Salomon}}]{schreck01PhysRevLett.87.080403}
\bibinfo{author}{\bibfnamefont{F.}~\bibnamefont{Schreck}},
  \bibinfo{author}{\bibfnamefont{L.}~\bibnamefont{Khaykovich}},
  \bibinfo{author}{\bibfnamefont{K.~L.} \bibnamefont{Corwin}},
  \bibinfo{author}{\bibfnamefont{G.}~\bibnamefont{Ferrari}},
  \bibinfo{author}{\bibfnamefont{T.}~\bibnamefont{Bourdel}},
  \bibinfo{author}{\bibfnamefont{J.}~\bibnamefont{Cubizolles}},
  \bibnamefont{and} \bibinfo{author}{\bibfnamefont{C.}~\bibnamefont{Salomon}},
  \bibinfo{journal}{Phys. Rev. Lett.} \textbf{\bibinfo{volume}{87}},
  \bibinfo{pages}{080403} (\bibinfo{year}{2001}).

\bibitem[{\citenamefont{Hadzibabic et~al.}(2002)\citenamefont{Hadzibabic, Stan,
  Dieckmann, Gupta, Zwierlein, G\"orlitz, and
  Ketterle}}]{hadzibabic02PhysRevLett.88.160401}
\bibinfo{author}{\bibfnamefont{Z.}~\bibnamefont{Hadzibabic}},
  \bibinfo{author}{\bibfnamefont{C.~A.} \bibnamefont{Stan}},
  \bibinfo{author}{\bibfnamefont{K.}~\bibnamefont{Dieckmann}},
  \bibinfo{author}{\bibfnamefont{S.}~\bibnamefont{Gupta}},
  \bibinfo{author}{\bibfnamefont{M.~W.} \bibnamefont{Zwierlein}},
  \bibinfo{author}{\bibfnamefont{A.}~\bibnamefont{G\"orlitz}},
  \bibnamefont{and} \bibinfo{author}{\bibfnamefont{W.}~\bibnamefont{Ketterle}},
  \bibinfo{journal}{Phys. Rev. Lett.} \textbf{\bibinfo{volume}{88}},
  \bibinfo{pages}{160401} (\bibinfo{year}{2002}).

\bibitem[{\citenamefont{Ferrari et~al.}(2002)\citenamefont{Ferrari, Inguscio,
  Jastrzebski, Modugno, Roati, and Simoni}}]{ferrari02PhysRevLett.89.053202}
\bibinfo{author}{\bibfnamefont{G.}~\bibnamefont{Ferrari}},
  \bibinfo{author}{\bibfnamefont{M.}~\bibnamefont{Inguscio}},
  \bibinfo{author}{\bibfnamefont{W.}~\bibnamefont{Jastrzebski}},
  \bibinfo{author}{\bibfnamefont{G.}~\bibnamefont{Modugno}},
  \bibinfo{author}{\bibfnamefont{G.}~\bibnamefont{Roati}}, \bibnamefont{and}
  \bibinfo{author}{\bibfnamefont{A.}~\bibnamefont{Simoni}},
  \bibinfo{journal}{Phys. Rev. Lett.} \textbf{\bibinfo{volume}{89}},
  \bibinfo{pages}{053202} (\bibinfo{year}{2002}).

\bibitem[{\citenamefont{Roati et~al.}(2002)\citenamefont{Roati, Riboli,
  Modugno, and Inguscio}}]{roati02PhysRevLett.89.150403}
\bibinfo{author}{\bibfnamefont{G.}~\bibnamefont{Roati}},
  \bibinfo{author}{\bibfnamefont{F.}~\bibnamefont{Riboli}},
  \bibinfo{author}{\bibfnamefont{G.}~\bibnamefont{Modugno}}, \bibnamefont{and}
  \bibinfo{author}{\bibfnamefont{M.}~\bibnamefont{Inguscio}},
  \bibinfo{journal}{Phys. Rev. Lett.} \textbf{\bibinfo{volume}{89}},
  \bibinfo{pages}{150403} (\bibinfo{year}{2002}).

\bibitem[{\citenamefont{Inouye et~al.}(2004)\citenamefont{Inouye, Goldwin,
  Olsen, Ticknor, Bohn, and Jin}}]{inouye04PhysRevLett.93.183201}
\bibinfo{author}{\bibfnamefont{S.}~\bibnamefont{Inouye}},
  \bibinfo{author}{\bibfnamefont{J.}~\bibnamefont{Goldwin}},
  \bibinfo{author}{\bibfnamefont{M.~L.} \bibnamefont{Olsen}},
  \bibinfo{author}{\bibfnamefont{C.}~\bibnamefont{Ticknor}},
  \bibinfo{author}{\bibfnamefont{J.~L.} \bibnamefont{Bohn}}, \bibnamefont{and}
  \bibinfo{author}{\bibfnamefont{D.~S.} \bibnamefont{Jin}},
  \bibinfo{journal}{Phys. Rev. Lett.} \textbf{\bibinfo{volume}{93}},
  \bibinfo{pages}{183201} (\bibinfo{year}{2004}).

\bibitem[{\citenamefont{Ferlaino et~al.}(2006)\citenamefont{Ferlaino, D'Errico,
  Roati, Zaccanti, Inguscio, Modugno, and
  Simoni}}]{ferlaino06PhysRevA.73.040702}
\bibinfo{author}{\bibfnamefont{F.}~\bibnamefont{Ferlaino}},
  \bibinfo{author}{\bibfnamefont{C.}~\bibnamefont{D'Errico}},
  \bibinfo{author}{\bibfnamefont{G.}~\bibnamefont{Roati}},
  \bibinfo{author}{\bibfnamefont{M.}~\bibnamefont{Zaccanti}},
  \bibinfo{author}{\bibfnamefont{M.}~\bibnamefont{Inguscio}},
  \bibinfo{author}{\bibfnamefont{G.}~\bibnamefont{Modugno}}, \bibnamefont{and}
  \bibinfo{author}{\bibfnamefont{A.}~\bibnamefont{Simoni}},
  \bibinfo{journal}{Phys. Rev. A} \textbf{\bibinfo{volume}{73}},
  \bibinfo{pages}{040702} (\bibinfo{year}{2006}).

\bibitem[{\citenamefont{Deh et~al.}(2008)\citenamefont{Deh, Marzok, Zimmermann,
  and Courteille}}]{deh08PhysRevA.77.010701}
\bibinfo{author}{\bibfnamefont{B.}~\bibnamefont{Deh}},
  \bibinfo{author}{\bibfnamefont{C.}~\bibnamefont{Marzok}},
  \bibinfo{author}{\bibfnamefont{C.}~\bibnamefont{Zimmermann}},
  \bibnamefont{and} \bibinfo{author}{\bibfnamefont{P.~W.}
  \bibnamefont{Courteille}}, \bibinfo{journal}{Phys. Rev. A}
  \textbf{\bibinfo{volume}{77}}, \bibinfo{pages}{010701}
  (\bibinfo{year}{2008}).

\bibitem[{\citenamefont{Park et~al.}(2012)\citenamefont{Park, Wu, Santiago,
  Tiecke, Will, Ahmadi, and Zwierlein}}]{park12PhysRevA.85.051602}
\bibinfo{author}{\bibfnamefont{J.~W.} \bibnamefont{Park}},
  \bibinfo{author}{\bibfnamefont{C.-H.} \bibnamefont{Wu}},
  \bibinfo{author}{\bibfnamefont{I.}~\bibnamefont{Santiago}},
  \bibinfo{author}{\bibfnamefont{T.~G.} \bibnamefont{Tiecke}},
  \bibinfo{author}{\bibfnamefont{S.}~\bibnamefont{Will}},
  \bibinfo{author}{\bibfnamefont{P.}~\bibnamefont{Ahmadi}}, \bibnamefont{and}
  \bibinfo{author}{\bibfnamefont{M.~W.} \bibnamefont{Zwierlein}},
  \bibinfo{journal}{Phys. Rev. A} \textbf{\bibinfo{volume}{85}},
  \bibinfo{pages}{051602} (\bibinfo{year}{2012}).

\bibitem[{\citenamefont{Ferrier-Barbut
  et~al.}(2014)\citenamefont{Ferrier-Barbut, Delehaye, Laurent, Grier, Pierce,
  Rem, Chevy, and Salomon}}]{ferrier-Barbut14Science.345.1035}
\bibinfo{author}{\bibfnamefont{I.}~\bibnamefont{Ferrier-Barbut}},
  \bibinfo{author}{\bibfnamefont{M.}~\bibnamefont{Delehaye}},
  \bibinfo{author}{\bibfnamefont{S.}~\bibnamefont{Laurent}},
  \bibinfo{author}{\bibfnamefont{A.~T.} \bibnamefont{Grier}},
  \bibinfo{author}{\bibfnamefont{M.}~\bibnamefont{Pierce}},
  \bibinfo{author}{\bibfnamefont{B.~S.} \bibnamefont{Rem}},
  \bibinfo{author}{\bibfnamefont{F.}~\bibnamefont{Chevy}}, \bibnamefont{and}
  \bibinfo{author}{\bibfnamefont{C.}~\bibnamefont{Salomon}},
  \bibinfo{journal}{Science} \textbf{\bibinfo{volume}{345}},
  \bibinfo{pages}{1035} (\bibinfo{year}{2014}).

\bibitem[{\citenamefont{Stuhler et~al.}(2005)\citenamefont{Stuhler, Griesmaier,
  Koch, Fattori, Pfau, Giovanazzi, Pedri, and
  Santos}}]{stuhler05PhysRevLett.95.150406}
\bibinfo{author}{\bibfnamefont{J.}~\bibnamefont{Stuhler}},
  \bibinfo{author}{\bibfnamefont{A.}~\bibnamefont{Griesmaier}},
  \bibinfo{author}{\bibfnamefont{T.}~\bibnamefont{Koch}},
  \bibinfo{author}{\bibfnamefont{M.}~\bibnamefont{Fattori}},
  \bibinfo{author}{\bibfnamefont{T.}~\bibnamefont{Pfau}},
  \bibinfo{author}{\bibfnamefont{S.}~\bibnamefont{Giovanazzi}},
  \bibinfo{author}{\bibfnamefont{P.}~\bibnamefont{Pedri}}, \bibnamefont{and}
  \bibinfo{author}{\bibfnamefont{L.}~\bibnamefont{Santos}},
  \bibinfo{journal}{Phys. Rev. Lett.} \textbf{\bibinfo{volume}{95}},
  \bibinfo{pages}{150406} (\bibinfo{year}{2005}).

\bibitem[{\citenamefont{Ospelkaus et~al.}(2008)\citenamefont{Ospelkaus,
  Pe{\'e}r, Ni, Zirbel, Neyenhuis, Kotochigova, Julienne, Ye, and
  Jin}}]{ospelkaus08NaturePhys.4.622}
\bibinfo{author}{\bibfnamefont{S.}~\bibnamefont{Ospelkaus}},
  \bibinfo{author}{\bibfnamefont{A.}~\bibnamefont{Pe{\'e}r}},
  \bibinfo{author}{\bibfnamefont{K.~K.} \bibnamefont{Ni}},
  \bibinfo{author}{\bibfnamefont{J.~J.} \bibnamefont{Zirbel}},
  \bibinfo{author}{\bibfnamefont{B.}~\bibnamefont{Neyenhuis}},
  \bibinfo{author}{\bibfnamefont{S.}~\bibnamefont{Kotochigova}},
  \bibinfo{author}{\bibfnamefont{P.~S.} \bibnamefont{Julienne}},
  \bibinfo{author}{\bibfnamefont{J.}~\bibnamefont{Ye}}, \bibnamefont{and}
  \bibinfo{author}{\bibfnamefont{D.~S.} \bibnamefont{Jin}},
  \bibinfo{journal}{Nature Phys.} \textbf{\bibinfo{volume}{4}},
  \bibinfo{pages}{622} (\bibinfo{year}{2008}).

\bibitem[{\citenamefont{Ni et~al.}(2008)\citenamefont{Ni, Ospelkaus,
  de~Miranda, Pe'er, Neyenhuis, Zirbel, Kotochigova, Julienne, Jin, and
  Ye}}]{ni08Science.322.5899}
\bibinfo{author}{\bibfnamefont{K.-K.} \bibnamefont{Ni}},
  \bibinfo{author}{\bibfnamefont{S.}~\bibnamefont{Ospelkaus}},
  \bibinfo{author}{\bibfnamefont{M.~H.~G.} \bibnamefont{de~Miranda}},
  \bibinfo{author}{\bibfnamefont{A.}~\bibnamefont{Pe'er}},
  \bibinfo{author}{\bibfnamefont{B.}~\bibnamefont{Neyenhuis}},
  \bibinfo{author}{\bibfnamefont{J.~J.} \bibnamefont{Zirbel}},
  \bibinfo{author}{\bibfnamefont{S.}~\bibnamefont{Kotochigova}},
  \bibinfo{author}{\bibfnamefont{P.~S.} \bibnamefont{Julienne}},
  \bibinfo{author}{\bibfnamefont{D.~S.} \bibnamefont{Jin}}, \bibnamefont{and}
  \bibinfo{author}{\bibfnamefont{J.}~\bibnamefont{Ye}},
  \bibinfo{journal}{Science} \textbf{\bibinfo{volume}{322}},
  \bibinfo{pages}{231} (\bibinfo{year}{2008}).

\bibitem[{\citenamefont{Taglieber et~al.}(2008)\citenamefont{Taglieber, Voigt,
  Aoki, H\"ansch, and Dieckmann}}]{taglieber08PhysRevLett.100.010401}
\bibinfo{author}{\bibfnamefont{M.}~\bibnamefont{Taglieber}},
  \bibinfo{author}{\bibfnamefont{A.-C.} \bibnamefont{Voigt}},
  \bibinfo{author}{\bibfnamefont{T.}~\bibnamefont{Aoki}},
  \bibinfo{author}{\bibfnamefont{T.~W.} \bibnamefont{H\"ansch}},
  \bibnamefont{and}
  \bibinfo{author}{\bibfnamefont{K.}~\bibnamefont{Dieckmann}},
  \bibinfo{journal}{Phys. Rev. Lett.} \textbf{\bibinfo{volume}{100}},
  \bibinfo{pages}{010401} (\bibinfo{year}{2008}).

\bibitem[{\citenamefont{Voigt et~al.}(2009)\citenamefont{Voigt, Taglieber,
  Costa, Aoki, Wieser, H\"ansch, and
  Dieckmann}}]{voigt09PhysRevLett.102.020405}
\bibinfo{author}{\bibfnamefont{A.-C.} \bibnamefont{Voigt}},
  \bibinfo{author}{\bibfnamefont{M.}~\bibnamefont{Taglieber}},
  \bibinfo{author}{\bibfnamefont{L.}~\bibnamefont{Costa}},
  \bibinfo{author}{\bibfnamefont{T.}~\bibnamefont{Aoki}},
  \bibinfo{author}{\bibfnamefont{W.}~\bibnamefont{Wieser}},
  \bibinfo{author}{\bibfnamefont{T.~W.} \bibnamefont{H\"ansch}},
  \bibnamefont{and}
  \bibinfo{author}{\bibfnamefont{K.}~\bibnamefont{Dieckmann}},
  \bibinfo{journal}{Phys. Rev. Lett.} \textbf{\bibinfo{volume}{102}},
  \bibinfo{pages}{020405} (\bibinfo{year}{2009}).

\bibitem[{\citenamefont{Pal et~al.}(2014)\citenamefont{Pal, Debatin, Gambari,
  Lam, Brachmann, and Dieckmann}}]{pal14DAMOPSessionB3.8}
\bibinfo{author}{\bibfnamefont{S.}~\bibnamefont{Pal}},
  \bibinfo{author}{\bibfnamefont{M.}~\bibnamefont{Debatin}},
  \bibinfo{author}{\bibfnamefont{J.}~\bibnamefont{Gambari}},
  \bibinfo{author}{\bibfnamefont{M.}~\bibnamefont{Lam}},
  \bibinfo{author}{\bibfnamefont{J.}~\bibnamefont{Brachmann}},
  \bibnamefont{and}
  \bibinfo{author}{\bibfnamefont{K.}~\bibnamefont{Dieckmann}},
  \bibinfo{journal}{http://meetings.aps.org/Meeting/DAMOP14/Session/B3.8}
  (\bibinfo{year}{2014}).

\bibitem[{\citenamefont{Park et~al.}(2014)\citenamefont{Park, Wu, Schloss,
  Wang, Will, and Zwierlein}}]{park14DAMOPSessionB3.6}
\bibinfo{author}{\bibfnamefont{J.~W.} \bibnamefont{Park}},
  \bibinfo{author}{\bibfnamefont{C.-H.} \bibnamefont{Wu}},
  \bibinfo{author}{\bibfnamefont{J.}~\bibnamefont{Schloss}},
  \bibinfo{author}{\bibfnamefont{Q.}~\bibnamefont{Wang}},
  \bibinfo{author}{\bibfnamefont{S.}~\bibnamefont{Will}}, \bibnamefont{and}
  \bibinfo{author}{\bibfnamefont{M.}~\bibnamefont{Zwierlein}},
  \bibinfo{journal}{http://meetings.aps.org/Meeting/DAMOP14/Session/B3.6}
  (\bibinfo{year}{2014}).

\bibitem[{\citenamefont{Vengalattore et~al.}(2008)\citenamefont{Vengalattore,
  Leslie, Guzman, and Stamper-Kurn}}]{vengalattore08PhysRevLett.100.170403}
\bibinfo{author}{\bibfnamefont{M.}~\bibnamefont{Vengalattore}},
  \bibinfo{author}{\bibfnamefont{S.~R.} \bibnamefont{Leslie}},
  \bibinfo{author}{\bibfnamefont{J.}~\bibnamefont{Guzman}}, \bibnamefont{and}
  \bibinfo{author}{\bibfnamefont{D.~M.} \bibnamefont{Stamper-Kurn}},
  \bibinfo{journal}{Phys. Rev. Lett.} \textbf{\bibinfo{volume}{100}},
  \bibinfo{pages}{170403} (\bibinfo{year}{2008}).

\bibitem[{\citenamefont{Ospelkaus et~al.}(2009)\citenamefont{Ospelkaus, Ni,
  de~Miranda, Neyenhuis, Wang, Kotochigova, Julienne, Jin, and
  Ye}}]{ospelkaus09FaradayDiscuss142.351}
\bibinfo{author}{\bibfnamefont{S.}~\bibnamefont{Ospelkaus}},
  \bibinfo{author}{\bibfnamefont{K.-K.} \bibnamefont{Ni}},
  \bibinfo{author}{\bibfnamefont{M.~H.~G.} \bibnamefont{de~Miranda}},
  \bibinfo{author}{\bibfnamefont{B.}~\bibnamefont{Neyenhuis}},
  \bibinfo{author}{\bibfnamefont{D.}~\bibnamefont{Wang}},
  \bibinfo{author}{\bibfnamefont{S.}~\bibnamefont{Kotochigova}},
  \bibinfo{author}{\bibfnamefont{P.~S.} \bibnamefont{Julienne}},
  \bibinfo{author}{\bibfnamefont{D.~S.} \bibnamefont{Jin}}, \bibnamefont{and}
  \bibinfo{author}{\bibfnamefont{J.}~\bibnamefont{Ye}},
  \bibinfo{journal}{Faraday Discuss.} \textbf{\bibinfo{volume}{142}},
  \bibinfo{pages}{351} (\bibinfo{year}{2009}).

\bibitem[{\citenamefont{Lu et~al.}(2011)\citenamefont{Lu, Burdick, Youn, and
  Lev}}]{lu11PhysRevLett.107.190401}
\bibinfo{author}{\bibfnamefont{M.}~\bibnamefont{Lu}},
  \bibinfo{author}{\bibfnamefont{N.~Q.} \bibnamefont{Burdick}},
  \bibinfo{author}{\bibfnamefont{S.~H.} \bibnamefont{Youn}}, \bibnamefont{and}
  \bibinfo{author}{\bibfnamefont{B.~L.} \bibnamefont{Lev}},
  \bibinfo{journal}{Phys. Rev. Lett.} \textbf{\bibinfo{volume}{107}},
  \bibinfo{pages}{190401} (\bibinfo{year}{2011}).

\bibitem[{\citenamefont{Aikawa et~al.}(2012)\citenamefont{Aikawa, Frisch, Mark,
  Baier, Rietzler, Grimm, and Ferlaino}}]{aikawa12PhysRevLett.108.210401}
\bibinfo{author}{\bibfnamefont{K.}~\bibnamefont{Aikawa}},
  \bibinfo{author}{\bibfnamefont{A.}~\bibnamefont{Frisch}},
  \bibinfo{author}{\bibfnamefont{M.}~\bibnamefont{Mark}},
  \bibinfo{author}{\bibfnamefont{S.}~\bibnamefont{Baier}},
  \bibinfo{author}{\bibfnamefont{A.}~\bibnamefont{Rietzler}},
  \bibinfo{author}{\bibfnamefont{R.}~\bibnamefont{Grimm}}, \bibnamefont{and}
  \bibinfo{author}{\bibfnamefont{F.}~\bibnamefont{Ferlaino}},
  \bibinfo{journal}{Phys. Rev. Lett.} \textbf{\bibinfo{volume}{108}},
  \bibinfo{pages}{210401} (\bibinfo{year}{2012}).

\bibitem[{\citenamefont{Wu et~al.}(2011)\citenamefont{Wu, Santiago, Park,
  Ahmadi, and Zwierlein}}]{wu11PhysRevA.84.011601}
\bibinfo{author}{\bibfnamefont{C.-H.} \bibnamefont{Wu}},
  \bibinfo{author}{\bibfnamefont{I.}~\bibnamefont{Santiago}},
  \bibinfo{author}{\bibfnamefont{J.~W.} \bibnamefont{Park}},
  \bibinfo{author}{\bibfnamefont{P.}~\bibnamefont{Ahmadi}}, \bibnamefont{and}
  \bibinfo{author}{\bibfnamefont{M.~W.} \bibnamefont{Zwierlein}},
  \bibinfo{journal}{Phys. Rev. A} \textbf{\bibinfo{volume}{84}},
  \bibinfo{pages}{011601} (\bibinfo{year}{2011}).

\bibitem[{\citenamefont{Shin et~al.}(2008)\citenamefont{Shin, Schirotzek,
  Schunck, and Ketterle}}]{shin08PhysRevLett.101.070404}
\bibinfo{author}{\bibfnamefont{Y.-I.} \bibnamefont{Shin}},
  \bibinfo{author}{\bibfnamefont{A.}~\bibnamefont{Schirotzek}},
  \bibinfo{author}{\bibfnamefont{C.~H.} \bibnamefont{Schunck}},
  \bibnamefont{and} \bibinfo{author}{\bibfnamefont{W.}~\bibnamefont{Ketterle}},
  \bibinfo{journal}{Phys. Rev. Lett.} \textbf{\bibinfo{volume}{101}},
  \bibinfo{pages}{070404} (\bibinfo{year}{2008}).

\bibitem[{\citenamefont{Santos et~al.}(2000)\citenamefont{Santos, Shlyapnikov,
  Zoller, and Lewenstein}}]{santos00PhysRevLett.85.1791}
\bibinfo{author}{\bibfnamefont{L.}~\bibnamefont{Santos}},
  \bibinfo{author}{\bibfnamefont{G.~V.} \bibnamefont{Shlyapnikov}},
  \bibinfo{author}{\bibfnamefont{P.}~\bibnamefont{Zoller}}, \bibnamefont{and}
  \bibinfo{author}{\bibfnamefont{M.}~\bibnamefont{Lewenstein}},
  \bibinfo{journal}{Phys. Rev. Lett.} \textbf{\bibinfo{volume}{85}},
  \bibinfo{pages}{1791} (\bibinfo{year}{2000}).

\bibitem[{\citenamefont{Yi and You}(2000)}]{yi00PhysRevA.61.041604}
\bibinfo{author}{\bibfnamefont{S.}~\bibnamefont{Yi}} \bibnamefont{and}
  \bibinfo{author}{\bibfnamefont{L.}~\bibnamefont{You}},
  \bibinfo{journal}{Phys. Rev. A} \textbf{\bibinfo{volume}{61}},
  \bibinfo{pages}{041604} (\bibinfo{year}{2000}).

\bibitem[{\citenamefont{Lewenstein et~al.}(2004)\citenamefont{Lewenstein,
  Santos, Baranov, and Fehrmann}}]{lewenstein04PhysRevLett.92.050401}
\bibinfo{author}{\bibfnamefont{M.}~\bibnamefont{Lewenstein}},
  \bibinfo{author}{\bibfnamefont{L.}~\bibnamefont{Santos}},
  \bibinfo{author}{\bibfnamefont{M.~A.} \bibnamefont{Baranov}},
  \bibnamefont{and} \bibinfo{author}{\bibfnamefont{H.}~\bibnamefont{Fehrmann}},
  \bibinfo{journal}{Phys. Rev. Lett.} \textbf{\bibinfo{volume}{92}},
  \bibinfo{pages}{050401} (\bibinfo{year}{2004}).

\bibitem[{\citenamefont{Sengupta et~al.}(2007)\citenamefont{Sengupta, Dupuis,
  and Majumdar}}]{sengupta07PhysRevA.75.063625}
\bibinfo{author}{\bibfnamefont{K.}~\bibnamefont{Sengupta}},
  \bibinfo{author}{\bibfnamefont{N.}~\bibnamefont{Dupuis}}, \bibnamefont{and}
  \bibinfo{author}{\bibfnamefont{P.}~\bibnamefont{Majumdar}},
  \bibinfo{journal}{Phys. Rev. A} \textbf{\bibinfo{volume}{75}},
  \bibinfo{pages}{063625} (\bibinfo{year}{2007}).

\bibitem[{\citenamefont{Sinha and Sengupta}(2009)}]{sinha09PhysRevB.79.115124}
\bibinfo{author}{\bibfnamefont{S.}~\bibnamefont{Sinha}} \bibnamefont{and}
  \bibinfo{author}{\bibfnamefont{K.}~\bibnamefont{Sengupta}},
  \bibinfo{journal}{Phys. Rev. B} \textbf{\bibinfo{volume}{79}},
  \bibinfo{pages}{115124} (\bibinfo{year}{2009}).

\bibitem[{\citenamefont{B\"uchler and
  Blatter}(2003)}]{buchler03PhysRevLett.91.130404}
\bibinfo{author}{\bibfnamefont{H.~P.} \bibnamefont{B\"uchler}}
  \bibnamefont{and} \bibinfo{author}{\bibfnamefont{G.}~\bibnamefont{Blatter}},
  \bibinfo{journal}{Phys. Rev. Lett.} \textbf{\bibinfo{volume}{91}},
  \bibinfo{pages}{130404} (\bibinfo{year}{2003}).

\bibitem[{\citenamefont{B\"uchler and
  Blatter}(2004)}]{buchler04PhysRevA.69.063603}
\bibinfo{author}{\bibfnamefont{H.~P.} \bibnamefont{B\"uchler}}
  \bibnamefont{and} \bibinfo{author}{\bibfnamefont{G.}~\bibnamefont{Blatter}},
  \bibinfo{journal}{Phys. Rev. A} \textbf{\bibinfo{volume}{69}},
  \bibinfo{pages}{063603} (\bibinfo{year}{2004}).

\bibitem[{\citenamefont{Orth et~al.}(2009)\citenamefont{Orth, Bergman, and
  Le~Hur}}]{orth09PhysRevA.80.023624}
\bibinfo{author}{\bibfnamefont{P.~P.} \bibnamefont{Orth}},
  \bibinfo{author}{\bibfnamefont{D.~L.} \bibnamefont{Bergman}},
  \bibnamefont{and} \bibinfo{author}{\bibfnamefont{K.}~\bibnamefont{Le~Hur}},
  \bibinfo{journal}{Phys. Rev. A} \textbf{\bibinfo{volume}{80}},
  \bibinfo{pages}{023624} (\bibinfo{year}{2009}).

\bibitem[{\citenamefont{Lim et~al.}(2010{\natexlab{a}})\citenamefont{Lim,
  Lazarides, Hemmerich, and Morais~Smith}}]{lim10PhysRevA.82.013616}
\bibinfo{author}{\bibfnamefont{L.-K.} \bibnamefont{Lim}},
  \bibinfo{author}{\bibfnamefont{A.}~\bibnamefont{Lazarides}},
  \bibinfo{author}{\bibfnamefont{A.}~\bibnamefont{Hemmerich}},
  \bibnamefont{and}
  \bibinfo{author}{\bibfnamefont{C.}~\bibnamefont{Morais~Smith}},
  \bibinfo{journal}{Phys. Rev. A} \textbf{\bibinfo{volume}{82}},
  \bibinfo{pages}{013616} (\bibinfo{year}{2010}{\natexlab{a}}).

\bibitem[{\citenamefont{Lim et~al.}(2010{\natexlab{b}})\citenamefont{Lim,
  Hemmerich, and Smith}}]{lim10PhysRevA.81.023404}
\bibinfo{author}{\bibfnamefont{L.-K.} \bibnamefont{Lim}},
  \bibinfo{author}{\bibfnamefont{A.}~\bibnamefont{Hemmerich}},
  \bibnamefont{and} \bibinfo{author}{\bibfnamefont{C.~M.} \bibnamefont{Smith}},
  \bibinfo{journal}{Phys. Rev. A} \textbf{\bibinfo{volume}{81}},
  \bibinfo{pages}{023404} (\bibinfo{year}{2010}{\natexlab{b}}).

\bibitem[{\citenamefont{Mathey et~al.}(2006)\citenamefont{Mathey, Tsai, and
  Neto}}]{mathey06PhysRevLett.97.030601}
\bibinfo{author}{\bibfnamefont{L.}~\bibnamefont{Mathey}},
  \bibinfo{author}{\bibfnamefont{S.-W.} \bibnamefont{Tsai}}, \bibnamefont{and}
  \bibinfo{author}{\bibfnamefont{A.~H.~C.} \bibnamefont{Neto}},
  \bibinfo{journal}{Phys. Rev. Lett.} \textbf{\bibinfo{volume}{97}},
  \bibinfo{pages}{030601} (\bibinfo{year}{2006}).

\bibitem[{\citenamefont{Klironomos and
  Tsai}(2007)}]{klironomos07PhysRevLett.99.100401}
\bibinfo{author}{\bibfnamefont{F.~D.} \bibnamefont{Klironomos}}
  \bibnamefont{and} \bibinfo{author}{\bibfnamefont{S.-W.} \bibnamefont{Tsai}},
  \bibinfo{journal}{Phys. Rev. Lett.} \textbf{\bibinfo{volume}{99}},
  \bibinfo{pages}{100401} (\bibinfo{year}{2007}).

\bibitem[{\citenamefont{Huang et~al.}(2013)\citenamefont{Huang, Lai, Shi, and
  Tsai}}]{huang13PhysRevB.88.054504}
\bibinfo{author}{\bibfnamefont{W.-M.} \bibnamefont{Huang}},
  \bibinfo{author}{\bibfnamefont{C.-Y.} \bibnamefont{Lai}},
  \bibinfo{author}{\bibfnamefont{C.}~\bibnamefont{Shi}}, \bibnamefont{and}
  \bibinfo{author}{\bibfnamefont{S.-W.} \bibnamefont{Tsai}},
  \bibinfo{journal}{Phys. Rev. B} \textbf{\bibinfo{volume}{88}},
  \bibinfo{pages}{054504} (\bibinfo{year}{2013}).

\bibitem[{\citenamefont{Bardeen et~al.}(1957)\citenamefont{Bardeen, Cooper, and
  Schrieffer}}]{bardeen57PhysRev.108.1175}
\bibinfo{author}{\bibfnamefont{J.}~\bibnamefont{Bardeen}},
  \bibinfo{author}{\bibfnamefont{L.~N.} \bibnamefont{Cooper}},
  \bibnamefont{and} \bibinfo{author}{\bibfnamefont{J.~R.}
  \bibnamefont{Schrieffer}}, \bibinfo{journal}{Phys. Rev.}
  \textbf{\bibinfo{volume}{108}}, \bibinfo{pages}{1175} (\bibinfo{year}{1957}).

\bibitem[{\citenamefont{Anderson and Morel}(1961)}]{anderson61PhysRev.123.1911}
\bibinfo{author}{\bibfnamefont{P.~W.} \bibnamefont{Anderson}} \bibnamefont{and}
  \bibinfo{author}{\bibfnamefont{P.}~\bibnamefont{Morel}},
  \bibinfo{journal}{Phys. Rev.} \textbf{\bibinfo{volume}{123}},
  \bibinfo{pages}{1911} (\bibinfo{year}{1961}).

\bibitem[{\citenamefont{Balian and Werthamer}(1963)}]{balian63PhysRev.131.1553}
\bibinfo{author}{\bibfnamefont{R.}~\bibnamefont{Balian}} \bibnamefont{and}
  \bibinfo{author}{\bibfnamefont{N.~R.} \bibnamefont{Werthamer}},
  \bibinfo{journal}{Phys. Rev.} \textbf{\bibinfo{volume}{131}},
  \bibinfo{pages}{1553} (\bibinfo{year}{1963}).

\bibitem[{\citenamefont{Osheroff et~al.}(1972)\citenamefont{Osheroff,
  Richardson, and Lee}}]{osheroff72PhysRevLett.28.885}
\bibinfo{author}{\bibfnamefont{D.~D.} \bibnamefont{Osheroff}},
  \bibinfo{author}{\bibfnamefont{R.~C.} \bibnamefont{Richardson}},
  \bibnamefont{and} \bibinfo{author}{\bibfnamefont{D.~M.} \bibnamefont{Lee}},
  \bibinfo{journal}{Phys. Rev. Lett.} \textbf{\bibinfo{volume}{28}},
  \bibinfo{pages}{885} (\bibinfo{year}{1972}).

\bibitem[{\citenamefont{Bednorz and M{\"u}ller}(1986)}]{bednorz86ZPhysB.64.189}
\bibinfo{author}{\bibfnamefont{J.~G.} \bibnamefont{Bednorz}} \bibnamefont{and}
  \bibinfo{author}{\bibfnamefont{K.~A.} \bibnamefont{M{\"u}ller}},
  \bibinfo{journal}{Z. Phys. B} \textbf{\bibinfo{volume}{64}},
  \bibinfo{pages}{189} (\bibinfo{year}{1986}).

\bibitem[{\citenamefont{Scalapino}(1995)}]{scalapino95PhysRep.250.329}
\bibinfo{author}{\bibfnamefont{D.}~\bibnamefont{Scalapino}},
  \bibinfo{journal}{Physics Reports} \textbf{\bibinfo{volume}{250}},
  \bibinfo{pages}{329 } (\bibinfo{year}{1995}), ISSN \bibinfo{issn}{0370-1573}.

\bibitem[{\citenamefont{Tsuei and Kirtley}(2000)}]{tsuei00RevModPhys.72.969}
\bibinfo{author}{\bibfnamefont{C.~C.} \bibnamefont{Tsuei}} \bibnamefont{and}
  \bibinfo{author}{\bibfnamefont{J.~R.} \bibnamefont{Kirtley}},
  \bibinfo{journal}{Rev. Mod. Phys.} \textbf{\bibinfo{volume}{72}},
  \bibinfo{pages}{969} (\bibinfo{year}{2000}).

\bibitem[{\citenamefont{Micnas et~al.}(1988)\citenamefont{Micnas, Ranninger,
  Robaszkiewicz, and Tabor}}]{micnas88PhysRevB.37.9410}
\bibinfo{author}{\bibfnamefont{R.}~\bibnamefont{Micnas}},
  \bibinfo{author}{\bibfnamefont{J.}~\bibnamefont{Ranninger}},
  \bibinfo{author}{\bibfnamefont{S.}~\bibnamefont{Robaszkiewicz}},
  \bibnamefont{and} \bibinfo{author}{\bibfnamefont{S.}~\bibnamefont{Tabor}},
  \bibinfo{journal}{Phys. Rev. B} \textbf{\bibinfo{volume}{37}},
  \bibinfo{pages}{9410} (\bibinfo{year}{1988}).

\bibitem[{\citenamefont{Micnas et~al.}(1990)\citenamefont{Micnas, Ranninger,
  and Robaszkiewicz}}]{micnas90RevModPhys.62.113}
\bibinfo{author}{\bibfnamefont{R.}~\bibnamefont{Micnas}},
  \bibinfo{author}{\bibfnamefont{J.}~\bibnamefont{Ranninger}},
  \bibnamefont{and}
  \bibinfo{author}{\bibfnamefont{S.}~\bibnamefont{Robaszkiewicz}},
  \bibinfo{journal}{Rev. Mod. Phys.} \textbf{\bibinfo{volume}{62}},
  \bibinfo{pages}{113} (\bibinfo{year}{1990}).

\bibitem[{\citenamefont{Bukov and Pollet}(2014)}]{bukov14PhysRevB.89.094502}
\bibinfo{author}{\bibfnamefont{M.}~\bibnamefont{Bukov}} \bibnamefont{and}
  \bibinfo{author}{\bibfnamefont{L.}~\bibnamefont{Pollet}},
  \bibinfo{journal}{Phys. Rev. B} \textbf{\bibinfo{volume}{89}},
  \bibinfo{pages}{094502} (\bibinfo{year}{2014}).

\bibitem{balents13}
L.\ Balents, private communication.

\bibitem[{\citenamefont{Petsas et~al.}(1994)\citenamefont{Petsas, Coates, and
  Grynberg}}]{petsas94PhysRevA.50.5173}
\bibinfo{author}{\bibfnamefont{K.~I.} \bibnamefont{Petsas}},
  \bibinfo{author}{\bibfnamefont{A.~B.} \bibnamefont{Coates}},
  \bibnamefont{and} \bibinfo{author}{\bibfnamefont{G.}~\bibnamefont{Grynberg}},
  \bibinfo{journal}{Phys. Rev. A} \textbf{\bibinfo{volume}{50}},
  \bibinfo{pages}{5173} (\bibinfo{year}{1994}).

\bibitem[{\citenamefont{Guidoni et~al.}(1997)\citenamefont{Guidoni, Trich\'e,
  Verkerk, and Grynberg}}]{guidoni97PhysRevLett.79.3363}
\bibinfo{author}{\bibfnamefont{L.}~\bibnamefont{Guidoni}},
  \bibinfo{author}{\bibfnamefont{C.}~\bibnamefont{Trich\'e}},
  \bibinfo{author}{\bibfnamefont{P.}~\bibnamefont{Verkerk}}, \bibnamefont{and}
  \bibinfo{author}{\bibfnamefont{G.}~\bibnamefont{Grynberg}},
  \bibinfo{journal}{Phys. Rev. Lett.} \textbf{\bibinfo{volume}{79}},
  \bibinfo{pages}{3363} (\bibinfo{year}{1997}).

\bibitem[{\citenamefont{Albus et~al.}(2003)\citenamefont{Albus, Illuminati, and
  Eisert}}]{albus03PhysRevA.68.023606}
\bibinfo{author}{\bibfnamefont{A.}~\bibnamefont{Albus}},
  \bibinfo{author}{\bibfnamefont{F.}~\bibnamefont{Illuminati}},
  \bibnamefont{and} \bibinfo{author}{\bibfnamefont{J.}~\bibnamefont{Eisert}},
  \bibinfo{journal}{Phys. Rev. A} \textbf{\bibinfo{volume}{68}},
  \bibinfo{pages}{023606} (\bibinfo{year}{2003}).

\bibitem[{\citenamefont{Titvinidze et~al.}(2009)\citenamefont{Titvinidze,
  Snoek, and Hofstetter}}]{titvinidze09PhysRevB.79.144506}
\bibinfo{author}{\bibfnamefont{I.}~\bibnamefont{Titvinidze}},
  \bibinfo{author}{\bibfnamefont{M.}~\bibnamefont{Snoek}}, \bibnamefont{and}
  \bibinfo{author}{\bibfnamefont{W.}~\bibnamefont{Hofstetter}},
  \bibinfo{journal}{Phys. Rev. B} \textbf{\bibinfo{volume}{79}},
  \bibinfo{pages}{144506} (\bibinfo{year}{2009}).

\bibitem[{\citenamefont{Lin et~al.}(2010)\citenamefont{Lin, Zhao, and
  Liu}}]{lin10PhysRevB.81.045115}
\bibinfo{author}{\bibfnamefont{C.}~\bibnamefont{Lin}},
  \bibinfo{author}{\bibfnamefont{E.}~\bibnamefont{Zhao}}, \bibnamefont{and}
  \bibinfo{author}{\bibfnamefont{W.~V.} \bibnamefont{Liu}},
  \bibinfo{journal}{Phys. Rev. B} \textbf{\bibinfo{volume}{81}},
  \bibinfo{pages}{045115} (\bibinfo{year}{2010}).

\bibitem[{\citenamefont{Menotti et~al.}(2007)\citenamefont{Menotti, Trefzger,
  and Lewenstein}}]{menotti07PhysRevLett.98.235301}
\bibinfo{author}{\bibfnamefont{C.}~\bibnamefont{Menotti}},
  \bibinfo{author}{\bibfnamefont{C.}~\bibnamefont{Trefzger}}, \bibnamefont{and}
  \bibinfo{author}{\bibfnamefont{M.}~\bibnamefont{Lewenstein}},
  \bibinfo{journal}{Phys. Rev. Lett.} \textbf{\bibinfo{volume}{98}},
  \bibinfo{pages}{235301} (\bibinfo{year}{2007}).

\bibitem[{\citenamefont{Titvinidze et~al.}(2008)\citenamefont{Titvinidze,
  Snoek, and Hofstetter}}]{titvinidze08PhysRevLett.100.100401}
\bibinfo{author}{\bibfnamefont{I.}~\bibnamefont{Titvinidze}},
  \bibinfo{author}{\bibfnamefont{M.}~\bibnamefont{Snoek}}, \bibnamefont{and}
  \bibinfo{author}{\bibfnamefont{W.}~\bibnamefont{Hofstetter}},
  \bibinfo{journal}{Phys. Rev. Lett.} \textbf{\bibinfo{volume}{100}},
  \bibinfo{pages}{100401} (\bibinfo{year}{2008}).

\bibitem[{\citenamefont{Danshita and S\'a~de
  Melo}(2009)}]{danshita09PhysRevLett.103.225301}
\bibinfo{author}{\bibfnamefont{I.}~\bibnamefont{Danshita}} \bibnamefont{and}
  \bibinfo{author}{\bibfnamefont{C.~A.~R.} \bibnamefont{S\'a~de Melo}},
  \bibinfo{journal}{Phys. Rev. Lett.} \textbf{\bibinfo{volume}{103}},
  \bibinfo{pages}{225301} (\bibinfo{year}{2009}).
  
\bibitem{gorkovSov61PhysJETP:13.1018}
L.\ P.\ Gorkov and T.\ K.\ Melik-Barkhudarov, So.\ Phys.\ JETP \textbf{13}, (1961).

\bibitem{heiselberg00PhysRevLett.85.2418}
H.\ Heiselberg, C.\ J.\ Pethick, H.\ Smith, and L.\ Viverit, Phys.\ Rev.\ Lett.\ \textbf{85}, 2418 (2000).

\bibitem{kohn65PhysRevLett.15.524}
W.\ Kohn and J.\ M.\ Luttinger, Phys.\ Rev.\ Lett.\ \textbf{15}, 524 (1965).

\bibitem{efremov02PhysRevB.65.134519}
D.\ V.\ Efremov and L.\ Viverit, Phys.\ Rev.\ B \textbf{65}, 134519 (2002).

\bibitem{galitski03PhysRevB.67.144520}
V.\ M.\ Galitski and S.\ Das Sarma, Phys.\ Rev.\ B \textbf{67}, 144520 (2003).

\bibitem{deGennes89Book}
P.\ G.\ de Gennes, ``Superconductivity of Metals and Alloys," Redwood City, California: Addison-Wesley (1989).

\bibitem{landau79Book}
L.\ Landau and E.\ Lifshitz, ``Statistical Physics," New York: Pergamon (1979).

\bibitem{schaffer14JPhysCondensMatter.26.423201}
A.\ M.\ Black-Schaffer and C.\ Honerkamp, J.\ Phys.: Condens.\ Matter \textbf{26}, 423201 (2014).

\bibitem{gilmore15}
R.\ Gilmore, private communication (2015).

\bibitem{abramowitz64}
M.\ Abramowitz and I.\ A.\ Stegun, ``Handbook of Mathematical Functions with Formulas, Graphs, and Mathematical Tables," National Bureau of Standards (1964).


\bibitem{gruner94DensityWavesInSolids}
G.\ Gr{\"u}ner, ``Density Waves in Solids," Cambridge, Massachusetts: Perseus Publishing (1994).

\bibitem{dzyaloshinskii88JETP.94.344}
I.\ E.\ Dzyaloshinskii and V.\ M.\ Yakovenko, Sov.\ Phys.\ JETP \textbf{67}, 844 (1988).

\bibitem{zheleznyak97PhysRevB.55.3200}
A.\ T.\ Zheleznyak, V.\ M.\ Yakovenko, and I.\ E.\ Dzyaloshinskii, Phys.\ Rev.\ B \textbf{55}, 3200 (1997).

\bibitem{stoof09UltracoldQuantumFieldsBook}
H.\ T.\ C.\ Stoof, K.\ B.\ Gubbels, D.\ B.\ M.\ Dickerscheid,
``Ultracold Quantum Fields," 
Dordrecht, The Netherlands: Springer (2009).

\end{thebibliography}

\end{document}